\pdfoutput=1
\documentclass[10pt, conference, compsocconf]{IEEEtran}

\usepackage{graphicx}
\usepackage{url}
\usepackage{listings}
\usepackage{caption}
\usepackage{subcaption}
\usepackage{paralist}
\usepackage{textcomp}
\usepackage{xspace}
\usepackage{ifthen}
\usepackage{amsmath}
\usepackage{amssymb}
\usepackage{color}
\usepackage{algpseudocode}
\usepackage{algorithm}
\usepackage[sort]{cite}
\usepackage{balance}

\begin{document}
%
\title{Counter-RAPTOR: Safeguarding Tor Against Active Routing Attacks}



%
\author{\IEEEauthorblockN{Yixin Sun,
Anne Edmundson,
Nick Feamster, 
Mung Chiang, 
Prateek Mittal}
\IEEEauthorblockA{Princeton University\\
\{yixins, annee, feamster, chiangm, pmittal\}@princeton.edu}}


\maketitle

\begin{abstract}
Tor is vulnerable to network-level adversaries who can observe both ends of the communication to deanonymize users. Recent work has shown that Tor is susceptible to the previously unknown active BGP routing attacks, called RAPTOR attacks, which expose Tor users to more network-level adversaries. In this paper, we aim to mitigate and detect such active routing attacks against Tor. First, we present a new measurement study on the resilience of the Tor network to active BGP prefix attacks. We show that ASes with high Tor bandwidth can be less resilient to attacks than other ASes. Second, we present a new Tor guard relay selection algorithm that incorporates resilience of relays into consideration to proactively mitigate such attacks. We show that the algorithm successfully improves the security for Tor clients by up to 36\% on average (up to 166\% for certain clients). Finally, we build a live BGP monitoring system that can detect routing anomalies on the Tor network in real time by performing an AS origin check and novel detection analytics. Our monitoring system successfully detects simulated attacks that are modeled after multiple known attack types as well as a \emph{real-world} hijack attack (performed by us), while having low false positive rates.
\end{abstract}


%

\section{Introduction}

The Tor network~\cite{dingledine2004tor} has been the most widely used system for anonymous communication that protects users' identities from untrusted parties who have access to user traffic. Tor serves millions of users and carries terabytes of traffic every day with its network of over 7,000 relays~\cite{tormetrics}. This makes Tor a popular target for adversaries who wish to compromise the anonymity of the users. 

Tor is vulnerable to traffic correlation attacks. An adversary who can observe the traffic at both ends of the communication path (i.e., between the Tor client and the entry relay, and between the exit relay and the destination server) can perform traffic analysis on packet sizes and timings to deanonymize the Tor users \cite{shmatikov2006timing, syverson2001towards}. Network-level adversaries, i.e., autonomous systems (ASes), that lie on the path between a Tor client and an entry relay, and between an exit relay and the destination server can deanonymize Tor clients \cite{edman2009awareness, feamster2004location, johnson2013users}. More recently, researchers have further exploited the dynamics of BGP routing to propose the new RAPTOR attacks~\cite{sun2015raptor}, which exaggerate this threat by enabling more network-level adversaries to be at a compromising position, including active BGP prefix attacks which were not previously studied on Tor. 

Building countermeasures to defend Tor against such malicious AS-level adversaries is an important challenge facing our community. Past work has explored AS-aware relay selection algorithms that minimize the chance of selecting Tor relays with the same AS lying on both ends of the communication paths \cite{akhoondi2012lastor, edman2009awareness, nithyanand2016measuring}. However, these works focus on mitigating \emph{passive} attacks in which AS-level adversaries only passively observe traffic instead of launching any \emph{active} attacks. These observations motivate our work on developing countermeasures against such active BGP attacks on Tor. 

First, we quantify the vulnerability of the current Tor network to active BGP prefix hijack and interception attacks. Second, we develop a novel Tor guard relay selection algorithm which incorporates AS resilience of Tor relays and proactively protects Tor clients from being affected by such attacks. Finally, we present a live BGP monitoring system on the Tor network that can detect routing anomalies in real time. To summarize, we make the following three contributions:




\textbf{Measurement on the Tor network.} In order to understand the importance of the threat and inspire defenses against the active routing attacks, we first measure the vulnerability of the current Tor network. 
Based on the current Internet topology ~\cite{topology} and Tor consensus data ~\cite{torconsensus}, we adapt an AS-resilience metric ~\cite{lad2007understanding} to measure resilience of the Tor network to BGP hijack attacks. Next, we develop a \emph{novel extension} of the metric to analyze interception attack scenarios and measure resilience to interception attacks launched by Tier 1 ASes. Our key findings are:

\begin{itemize}
\item Some ASes that have high Tor bandwidth have low resiliences to hijack attacks, e.g., AS 16276 (OVH), which contains 339 Tor relays and only has a resilience value of 0.408 on a scale of $[0,1]$, indicating that in a hijack event, the probability of a Tor client (who uses relays in this AS) being deceived is close to 60\%. The cause of this lies in the topological features of the ASes in the AS hierarchy.
\item ASes have higher resiliences to interception attacks. However, some ASes (e.g., OVH) with high Tor relay bandwidth still have relatively low resilience (e.g., OVH has resilience 0.56 while the average is 0.8).
\item AS resilience varies depending on client location, and has high heterogeneity. For instance, OVH has resilience $>= 0.8$ for 20\% of the client ASes, while $<= 0.3$ for another 20\% of the client ASes.
\end{itemize}
\textbf{Proactive approach against active BGP attacks.} 
We propose and implement a novel Tor guard relay selection algorithm, which considers a resilience metric for ASes and protects the connection between Tor clients and Tor guard relay. \emph{Our guard relay selection algorithm is the first algorithm to incorporate resilience to active BGP routing attacks on Tor \cite{sun2015raptor} and the first countermeasure to proactively protect Tor clients from being affected by such attacks.} The algorithm combines resilience and bandwidth into relay selection to ensure security as well as performance. 
Our evaluation shows that the algorithm achieves up to 36\% improvement on average (up to 166\% for certain clients) in probability of Tor clients being resilient to a prefix hijack attack on guard relay and improves the anonymity bounds (computed by MATor \cite{backes2014nothing}) compared to the current Tor relay selection algorithm. At the same time, it only suffers minimal performance loss based on 
a large-scale evaluation on the Shadow emulator. 
\\
\textbf{Reactive approach against active BGP attacks.} To complement our proactive defense, we build and deploy a live monitoring system that monitors routing activities for Tor relays in real time. The monitoring system uses the real-time BGP routing information in addition to novel analytics-based hijack detection methods.  The system collects live BGP updates from BGP Stream \cite{bgpstream}, as well as the latest hourly Tor consensus data, and detects suspicious prefix announcements (affecting the Tor network) by performing AS origin check and analytics in real time. Our evaluation shows that most BGP updates that involve a Tor relay are only announced by a single AS (across all updates).  
Our detection analytics have a low false positives rate of 0.19\% on average. 
We also show that both the live AS origin check and the analytics successfully detected simulated attacks that are modeled after multiple known attack types as well as a \emph{real-world} BGP hijack attack (performed by us). The monitoring system will help enhance the transparency of the Tor network with regards to active BGP attacks. 

The paper is organized as follows. Section 2 provides a brief overview of background and related work on Tor.  Section 3 describes the metrics used to measure the resilience of Tor network to active BGP hijack/interception attacks and presents the results. Section 4 presents our new Tor guard relay selection algorithm and evaluation. Section 5 shows our design for the live monitoring system and describes our deployment experience.  Section 6 provides discussions on the current approaches and directions for future work. Finally, we conclude in Section 7.

\section{Background and Related Work}
Here we discuss network-level adversaries on the Tor network and past work on defending against such network-level adversaries. 

\subsection{Network Adversaries on Tor}
%
Feamster and Dingledine \cite{feamster2004location} first investigated AS-level adversaries in anonymity networks, and they showed that some ASes could appear on nearly 30\% of entry-exit pairs, although the anonymity networks have grown significantly since then. Murdoch and Zielinski \cite{murdoch2007sampled} later demonstrated the threat posed by network-level adversaries who can deanonymize users by performing traffic analysis. Furthermore, Edman and Syverson \cite{edman2009awareness} demonstrated that even given the explosive growth of Tor during the past years, still about 18\% of Tor circuits result in a single AS being able to observe both ends of the communication path. In 2013, Johnson \emph{et al.} \cite{johnson2013users} evaluated the security of Tor users over a period of time, and the results indicated that a network-level adversary with just low-bandwidth cost/budget could deanonymize any user within three months with over 50\% probability and within six months with over 80\% probability.

While all prior research, to our knowledge, focuses on passive adversaries, more recently, Sun \emph{et al.} \cite{sun2015raptor} proposed a new suite of attacks, called RAPTOR attacks, that discovered the threat posed by active AS-level adversaries who can perform active BGP routing attacks to put themselves onto the path between client-entry and/or exit-destination. The paper also showed that these routing attacks have occurred on the Tor network.  Using past BGP data, they demonstrated that Tor relays were affected in prefix hijack attacks. For example, in the Indosat hijack in 2014 \cite{indosat2014}, among the victim prefixes there were 44 Tor relays, and 33 of them were guard relays which had direct connections with Tor clients. 

\subsection{Defenses against Network Adversaries}
The existence of network-level adversaries motivates the research on AS-awareness in path selection in Tor. In 2012, Akhoondi \emph{et al.} \cite{akhoondi2012lastor} proposed LASTor, a Tor client that takes into account AS-level path and relay locations in selecting a path; our work differs by considering relays' resilience to active attacks and relays' capacity. Recently, Nithyanand \emph{et al.} \cite{nithyanand2016measuring} constructed a new Tor client, Astoria, which adopted a new path selection algorithm that considered more aspects - relay capacity, asymmetric routing, and colluding ASes. Barton \emph{et al.} \cite{barton2016denasa} also proposed a destination-naive AS-aware path selection, DeNASA, which avoided a list of suspect ASes when constructing circuits in advance. However, both Astoria and DeNASA only considers passive AS-level attackers and does not consider the case of active routing attacks.  

Most recently, Tan \emph{et al.} \cite{tan2016data} proposed a data-plane detection approach that periodically runs traceroute to detect longest-prefix attacks and update Tor relay descriptors upon anomaly detection, so that Tor users can pick guard relays correspondingly. Unfortunately, this approach cannot proactively protect Tor users who have already established connections with Tor guard relays which are under attack, and furthermore, the detection is not in real time - the periodic nature of traceroutes and hourly update of Tor consensus will both lead to delays in detection while the attacks may be short-lived (Sun \emph{et al.}~\cite{sun2015raptor} show that deanonymization accuracy can reach 90\% by performing a longest-prefix attack for less than 5 minutes). Thus, these observations motivate our work on developing countermeasures that can \emph{proactively} defend Tor users against active routing attacks, as well as a live monitoring system that can detect attacks on Tor in real time. 



\subsection{Resilience to Active Routing Attacks}


Lad \emph{et al.} \cite{lad2007understanding} investigated the relationship between Internet topology and prefix hijacking, and provided a metric for evaluating AS resilience to active prefix hijack attacks. While the study provides a foundational starting point for our work, it was conducted in 2007 when there were far fewer ASes than today and it only simulated a partial attack scenario of 1000 randomly selected ASes as attackers. Furthermore, it is not specific to the Tor network. In comparison to this work, we first adapt the metric to measure resilience of Tor to active hijack attacks, considering \emph{all} attack scenarios as well as \emph{weighted} attack scenarios based on top Tor client locations. We then devise a \emph{novel extension} to evaluate resilience to active \emph{interception} attacks. In Section \ref{sec:relayselection}, we incorporate the AS resilience metric into the Tor guard relay selection algorithm.

%
%
%

\section{Measuring Tor's Current State of Resilience to BGP Hijack and Interception Attacks}
\label{hijack_interception_measurement}

Network-level adversaries can launch BGP routing attacks by announcing BGP prefixes that they do not own. Over the past years, several well known attacks that affected large portions of the Internet \cite{indiahijack,syriahijack,many_hijack,mada_hijack,indosat2014} continuously show us the high vulnerability of BGP. The Tor network is no exception -- more than 90\% of BGP prefixes hosting Tor relays have prefix length shorter than /24, making them vulnerable to more-specific prefix attacks \cite{sun2015raptor}. However, the prevalence of equally-specific attacks and how these attacks affect Tor relays have not been well studied. This type of hijack attack tends to be more stealthy because unlike more-specific prefix attacks, equally-specific attacks may not be seen by all vantage points.  Furthermore, interception attacks which are more relevant for traffic analysis can only be equally-specific attacks, which we will explain in Section \ref{interception_methodology}. First, we show in a case study that the Tor network has been affected by equally-specific attacks in past real-world hijacks. Second, we look at how to evaluate the Tor network in terms of susceptibility to equally-specific hijack and interception attacks. These steps help quantify how vulnerable the Tor network is to real world network-level adversaries and also provide insights for developing countermeasures. 

\subsection{Motivation: Equally-Specific Prefix Attacks on Tor}
\label{subsec:equal_case_study}
Tor relays have already been affected in past known BGP attacks \cite{sun2015raptor}. As a motivating example, we consider the Indosat 2014 hijack \cite{indosat2014}, which affected the most number of Tor relays among all the attacks. The 44 Tor relays that were hijacked belonged to 23 prefixes. We found that \emph{all} these 23 prefixes were announced by the false origin, Indosat (ASN 4761), in the \emph{same length} as were announced by their true origin ASes. Table ~\ref{tbl:indosat} shows the prefix lengths of the 23 prefixes. We can see that equally-specific attacks are a real threat. Note that equally-specific attack \emph{is} a shorter path attack, as the traffic will go to the false origin AS only when its path is shorter (or, more preferred) than the path to the true origin AS.

\begin{table}[ht!]
\begin{center}
\small
  \begin{tabular}{ l  c  c  c c c c c c c}
    \hline
    Prefix Length & 13 & 14 & 16 & 18 & 19 & 20 & 21 & 23 & 24 \\ \hline
    \# of Prefixes & 3 & 1 & 3 & 1 & 1 & 2 & 5 & 2 & 5 \\
    \hline
  \end{tabular}
\end{center}
\caption{Prefix lengths of prefixes that cover Tor relays, and were affected in Indosat 2014 hijack. All of these prefixes were attacked by equally-specific prefix hijacks.}
\label{tbl:indosat}
\end{table}

\subsection{Resilience to prefix hijack attacks}
\label{hijack_methodology}

When an adversary announces an equally-specific BGP prefix to hijack the traffic, some ASes would be deceived by the false announcement and thus send traffic to the false origin AS (adversary) instead of the true origin AS. Previous work has studied how many, and which, ASes are affected by prefix hijack attacks using simulations of the entire Internet \cite{lad2007understanding}. We build off this work by adapting the AS resilience metric to the Tor network, by considering \emph{all} attack scenarios as well as \emph{weighted} attack scenarios based on top Tor client locations. 

\subsubsection{Definition of resilience}
\label{subsec:def_resil}

We first define the term \emph{resilience}. A source AS $v$ is \emph{resilient} to a hijack attack launched by a false origin AS $a$ on a true origin AS $t$, if $v$ is \emph{not deceived} by $a$ and still sends its traffic to $t$. 

\begin{itemize}
\item The \emph{origin-source-attacker resilience} of a particular true origin AS $t$ with a given source AS $v$ and attacking AS $a$, is defined as the \emph{probability} of $v$ being resilient and thus is always in the interval of $[0, 1]$. For instance, an origin-source-attacker resilience value of $0.4$ indicates 40\% probability of $v$ being resilient to the attack (or, 60\% probability of being deceived).
\item The \emph{origin-source resilience} of a particular true origin AS $t$ with a given source AS $v$ is defined as the probability of $v$ being resilient if \emph{any other ASes} launch a prefix attack on $t$.
\item The \emph{origin resilience} of a particular true origin AS $t$, with a given target set of source ASes (e.g., \emph{all} existing ASes or top ASes containing Tor clients), is defined as the averaged probability of the source ASes in the target set being resilient if \emph{any other ASes} launch a prefix attack on $t$.
\end{itemize}

In the following sub-sections, we show the detailed steps to calculate such resilience values of the ASes in the Tor network. 

\subsubsection{AS path prediction}
\label{subsec:path_predict}

In order to compute the probability of a source AS being resilient/deceived, we first need to predict its AS-level paths to both the false origin AS and true origin AS of destination. Gao \emph{et al.} have shown that AS level paths are determined mainly based on two preferences ~\cite{gao2001stable}: (1) Local Preference: customer route is preferred over peer route, which is preferred over provider route; (2) Shortest Path: Among paths with the highest local preference, paths with the shortest hops will be preferred.


Furthermore, the AS paths should also have the \emph{valley free} property ~\cite{gao2001inferring}. Thus, we use breadth first search to traverse the graph from a given source node based on this property and the preferences. We first explore provider-customer paths, which are the most preferred; next, we explore one peer-to-peer path followed by a sequence of provider-customer paths, which are the next preferred; finally, we explore customer-provider paths followed by an optional peer-to-peer path and then followed by a sequence of provider-customer paths. Nodes are explored in the most preferred to least preferred order, and those which are explored in the same step are equally preferred. This ordering will help accelerate the resilience calculation.

\subsubsection{Origin-source-attacker resilience for given $(t, v, a)$}
\label{subsec:single_resil}

Next, given the AS-level paths, we will then compute the probability of a source AS $v$ being resilient to the attack. If the best path from the source AS $v$ to the true origin $t$ is \emph{more preferred} than the best path to the false origin $a$, then the resilience would be 1; in the opposite case when it is \emph{less preferred}, the resilience would be 0. If they're equally preferred, the probability will be computed as follows: 

\begin{equation}
\bar{\beta}(t, v, a) = \frac {p(v,t)} {p(v,t) + p(v,a)}
\end{equation}

where $p(v,a)$ is the number of equally preferred paths from node $v$ to false origin $a$ and $p(v,t)$ is the number of equally preferred paths from node $v$ to true origin $t$.  

\subsubsection{Origin resilience}
\label{subsec:total_resil}

With the origin-source-attacker resiliences of each $(t, v, a)$ tuple, we can first compute the origin-source resilience of an origin AS $t$  for each source AS $v$ in the target set, by aggregating the origin-source-attacker resiliences of \emph{all attacking ASes} (representing \emph{all} attack scenarios which can be launched by \emph{any} AS $a$). Then, we can compute origin resilience of the origin AS $t$ by further aggregating the origin-source resiliences of the source ASes in the target set. The following equation illustrates the resilience computation when the target set of source ASes equals to the set of all ASes. 


\begin{equation}
R(t) = \sum_{a \in {\cal N}} \sum_{v \in {\cal N}} \frac {\bar{\beta}(t, v, a)} {(N-1)(N-2)}
\end{equation}

where ${\cal N}$ is the set of all ASes and $N$ is the total number of ASes.

We adapt the above metric to Tor by measuring the origin resilience of each AS that contains at least one Tor relay. Algorithm \ref{algo:calcres} shows the detailed steps to calculate the origin-source resilience for each Tor-related AS $t$ from a given source/client AS $v$. 
The origin resilience for each Tor-related AS $t$ can then be computed by aggregating over the target set of source/client ASes. In Section \ref{hijack_results}, we illustrate the results when (1) the target set is the set of all ASes, and (2) the target set is the set of top 95 Tor client ASes \cite{johnson2013users}. 




\begin{algorithm}
\caption{Origin-source resilience to hijacks for Tor-related ASes from a source AS \emph{v}.}
\label{algo:calcres}
\small
\begin{algorithmic}
\Function{CalcHijackResilience}{graph $G$, node $v$}
    \State \Call{CalcPathsFromNode}{$G,v$}
    \State $R[t] = 0$ $\forall$ Tor AS $t$
    \For{each reachable node $i$ from node $v$} 
    	\If{node $i$ contains Tor relays}
		\State $n \gets $ num. of nodes with less preferred paths than node $i$
		\State $R[i] \gets n + \sum_{a \in {\cal A}} \bar{\beta}(i,v,a)$ where ${\cal A}$ is the set of nodes with equally-preferred paths as node $i$
	\EndIf
    \EndFor
    \State $N \gets$ num. of nodes in $G$
    \State \Return $[R[i] / (N-2)$ for each node $i$ in $R]$
\EndFunction
\end{algorithmic}
\end{algorithm}



\subsection{Resilience to prefix interception attacks}
\label{interception_methodology}
Next, we derive a \emph{new} extension of the metric to measure the resilience of Tor-related ASes to prefix interception attacks. Launching a prefix interception attack requires one further step than prefix hijack attacks - the false origin AS needs to forward the hijacked traffic back to the true origin AS. Note that interception attacks can only happen with equally-specific prefix attacks; otherwise, if it's a more-specific prefix attack, the whole internet would be affected and the false origin AS would not be able to route the traffic back to the true origin. Prior work \cite{ballani2007study, pilosov2008stealing} has pointed out that to be able to do this, the false origin AS needs to satisfy a \emph{safety condition}: none of the ASes along the existing route from false origin AS to true origin AS should choose the invalid route advertised by the false origin AS, and thus the false origin AS can still use its existing route to forward the hijacked traffic back to the true origin. Thus, when making the invalid route announcement, there are two cases to consider: (1) if the false origin AS's existing route to the true origin AS is through a peer or customer route, then it's safe to make the false announcement to all its neighbors without affecting its existing route to the true origin; (2) if the false origin AS's existing route to the true origin AS is through a provider route, then it can only make the false announcement to its peers and customers, but not providers. 

Based on the above property, we modify Algorithm ~\ref{algo:calcres} to the following Algorithm~\ref{algo:calcintercept} to evaluate resilience to interception attacks. 

\begin{algorithm}
\caption{Origin-source resilience to interceptions for Tor-related ASes from a source AS \emph{v}.}
\label{algo:calcintercept}
\small
\begin{algorithmic}
\Function{CalcInterceptResilience}{graph $G$, node $v$}
    \State \Call{CalcPathsFromNode}{$G,v$}
    \State $R[t] = 0$ $\forall$ Tor AS $t$
    \For{each reachable node $i$ from node $v$} 
    	\If{node $i$ contains Tor relays}
		\State $n \gets $ num. of less preferred nodes than node $i$
		\State ${\cal N} \gets$ set of more preferred nodes than node $i$
		\If{existing route $v$ to $i$ is provider route}
			\State ${\cal N} \gets {\cal N} \cap {\cal M}$ where ${\cal M}$ contains all nodes $m$ for which $v$ to $m$ is provider route
		\EndIf
		\State ${\cal A} \gets$ set of equally preferred nodes as node $i$
		\If{existing route $v$ to $i$ is provider route}
			\State ${\cal P} \gets {\cal A} \cap {\cal M}$ where ${\cal M}$ contains all nodes $m$ for which $v$ to $m$ is provider route
			\State ${\cal A} \gets {\cal A} - {\cal P}$ 
		\EndIf
		\State $R[i] \gets n + len({\cal N}) + len({\cal P}) + \sum_{a \in {\cal A}} \bar{\beta}(i,v,a)$
	\EndIf
    \EndFor
    \State $N \gets$ num. of nodes in $G$
    \State \Return $[R[i] / (N-2)$ for each node $i$ in $R]$
\EndFunction
\end{algorithmic}
\end{algorithm}


\subsection{Hijack Resilience Results}
\label{hijack_results}

We obtained the list of Tor relays from the Tor consensus data on January 1, 2016 and retrieved their corresponding ASes. Then, we downloaded the AS topology published by CAIDA in January 2016. The AS topology contains 52,838 ASes, in which 1,185 ASes contain a total of 6,942 Tor relays. We first considered \emph{all} possible hijacking scenarios (any AS can be a potential attacker and the Tor client can be located in any AS) against each of the 1,185 Tor-related ASes, totaling $52,837 \times 1,185 = 62,611,845$ prefix hijacks. We used the methods described in Section \ref{hijack_methodology} to evaluate the origin resilience of each Tor-related AS. 

Since Tor clients are not evenly distributed across all ASes on the Internet, so we also consider and evaluate the resilience of only ASes that contain Tor guard relays from 95 top Tor client ASes \cite{johnson2013users} as the target set of source ASes. Furthermore, since Tor clients select relays in a bandwidth-weighted manner, so we also evaluate resilience weighted by Tor bandwidth of the ASes. 

\begin{figure}[ht!]
\centering
\includegraphics[width=.5\textwidth]{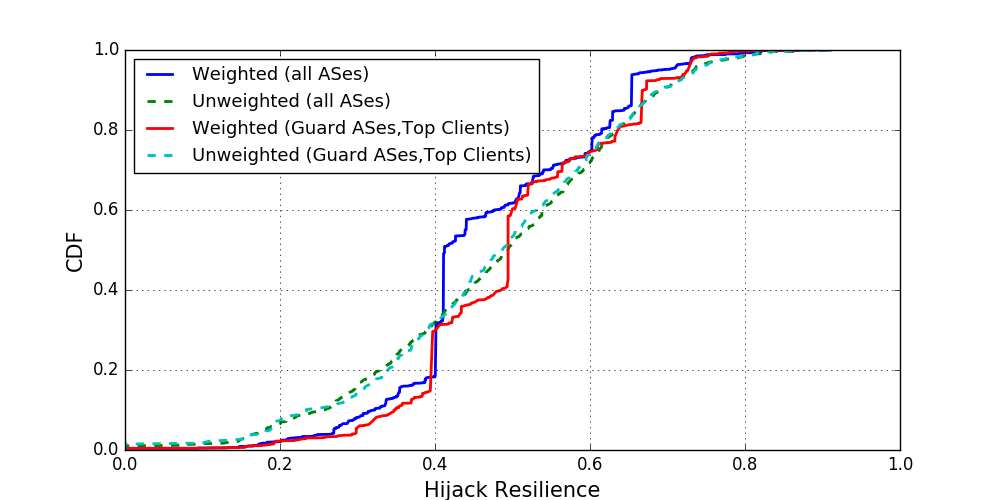}
\caption{Hijack Resilience for Tor-related ASes.}
\label{fig:hijack_as}
\end{figure}

Figure \ref{fig:hijack_as} shows the CDFs of AS resilience distribution for: (i) all Tor-related ASes with target set of source ASes being all ASes; (ii) all Tor-related ASes weighted by the accumulative Tor bandwidth of each AS, with target set being all ASes; (iii) all Tor Guard ASes with target set being top 95 Tor client ASes; (iv) all Tor Guard ASes weighted by bandwidth, with target set being top 95 Tor client ASes. 



\begin{figure}[ht!]
\centering
\includegraphics[width=.5\textwidth]{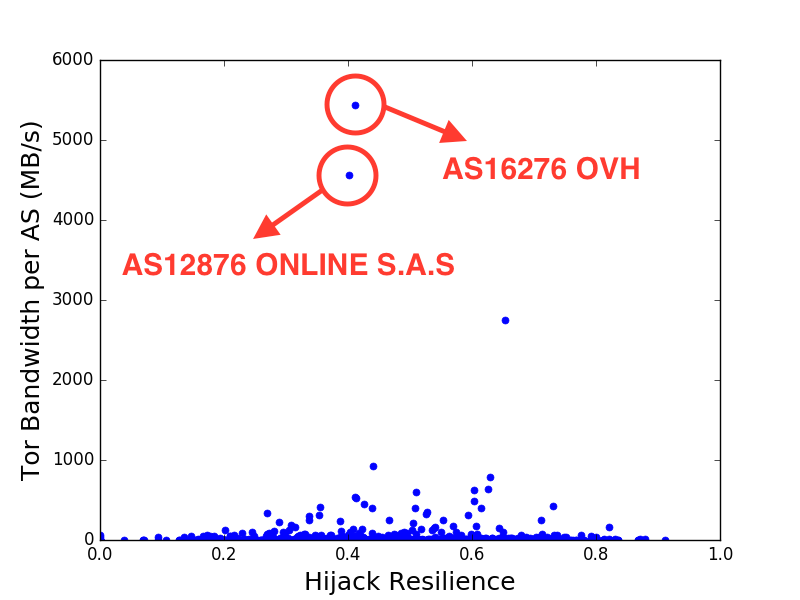}
\caption{Hijack Resilience and Corresponding Bandwidth per AS. OVH and S.A.S are clear outliers.}
\label{fig:ovh_bandwidth}
\end{figure}

{\bf Interpreting the hijack resilience.} All the four curves show high heterogeneity among AS resiliences. Let us first consider the two unweighted curves for all Tor ASes and only Guard ASes. They have almost identical distribution. Among them, about 20\% of the ASes have high resiliences $> 0.61$, indicating that in a hijack event, the averaged probability of a Tor client (who uses relays in the ASes) being deceived is smaller than $39\%$. However, there are also 20\% of the ASes with low resiliences $< 0.32$. The two weighted curves show some differences: they both have a steep jump at roughly 20\% point first, and then another jump at 30\% and 40\% points, respectively. This is due to two high-bandwidth ASes: i) AS12876 (ONLINE S.A.S), with resilience 0.4 from all source ASes (corresponding to the 20\% point in the all-ASes curve) and resilience 0.39 from top Tor client ASes (corresponding to the 20\% point in the top-client curve); ii) AS16276 (OVH), with resilience 0.41 from all source ASes (corresponding to the 30\% point in the all-ASes curve) and resilience 0.49 from top Tor client ASes (corresponding to the 40\% point in the top-client curve). Figure \ref{fig:ovh_bandwidth} plots resilience versus bandwidth per Tor AS. We can see clearly the two outliers OVH and ONLINE S.A.S. - high bandwidth, yet low resilience. 

The results lead to two question we want to answer: (i) why do some ASes (e.g., OVH) have relatively lower resiliences than others?, and (ii) since the origin resilience for an AS represents the \emph{averaged} probability across its clients of being resilient to hijacks, can some of its clients still have high origin-source resiliences even though the origin resilience is low?

\begin{figure}[ht!]
\centering
\includegraphics[width=.5\textwidth]{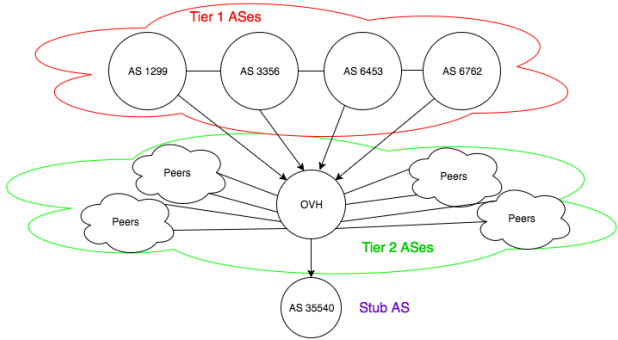}
\caption{Partial AS Graph containing OVH.}
\label{fig:ovh}
\end{figure}

{\bf Analyzing low resilience values of ASes.} To answer the first question, we choose AS16276 (OVH) as an example and conduct a deeper analysis on it. OVH's relatively lower resilience is due to its topological features in the AS hierarchy, shown in Figure \ref{fig:ovh}. OVH has 4 provider ASes (which are all tier-1 ASes), and it also has quite a number of peer ASes, while only having one customer AS (which is a stub AS). 
Since OVH only has one customer (AS 35540), so this lone customer AS is the only AS that is guaranteed to remain unaffected in a prefix hijack attack (while for other ASes, the impact depends on their location). For instance, if the hijacking AS is the customer of a peer AS of OVH, then all other customers of the particular peer AS will be affected, and even customers of other peer ASes may also stand a chance of being affected, while the four Tier 1 ASes will stay unaffected. Thus, if a Tor client is located in AS 35540 (OVH's only customer), then selecting a Tor guard relay in OVH will largely eliminate the chance of being affected by attacks on OVH and result in perfect resilience of value 1. 


\begin{figure*}[ht!]
\def\arraystretch{1.1}
\centering\captionsetup{aboveskip=6pt}
\begin{subfigure}{.5\textwidth}
  \centering\captionsetup{width=.9\linewidth}%
  \includegraphics[width=\linewidth]{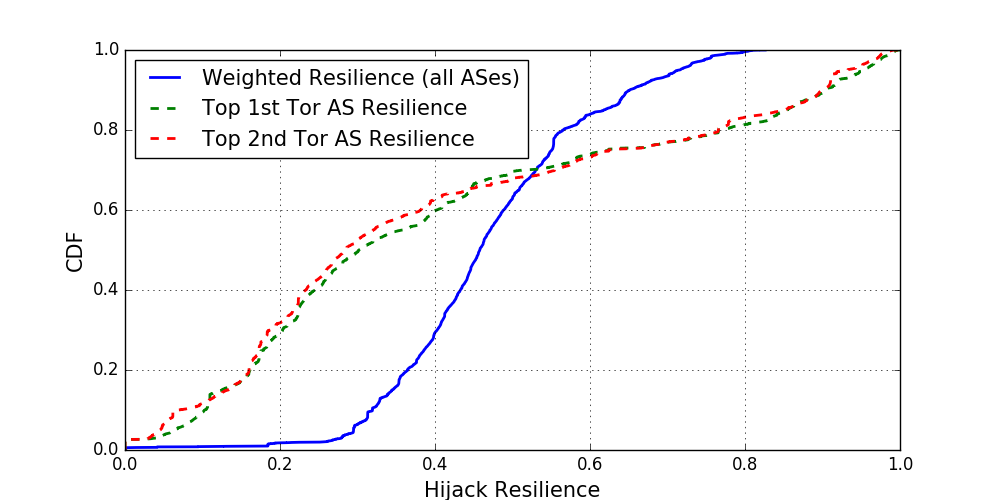}
  \caption{Hijack Resilience from All Source ASes}
  \label{fig:all_hijack}
\end{subfigure}%
\begin{subfigure}{.5\textwidth}
  \centering\captionsetup{width=.9\linewidth}%
  \includegraphics[width=\linewidth]{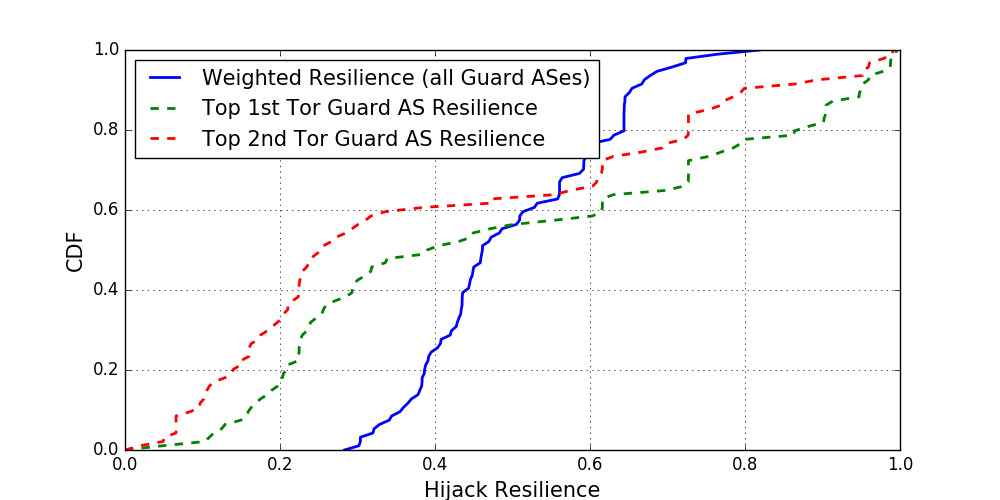}
  \caption{Hijack Resilience from Top Client ASes}
  \label{fig:topclient_hijack}
\end{subfigure}
\caption{Hijack Resiliences for Top-Bandwidth ASes across Different Clients.}
\label{fig:hijack_client}
\end{figure*}

{\bf Heterogeneity in resilience across clients.} To answer the second question, we plot the CDF of origin-source resilience distribution of each source/client AS for the top 2 ASes with highest Tor bandwidth (AS16276 and AS12876), as shown in Figure \ref{fig:all_hijack}. The weighted resilience curve represents the origin-source resilience for each source AS averaged over all Tor ASes (weighted by each AS' Tor bandwidth). Both AS17276 and AS12876 exhibit significantly higher heterogeneity compared to the weighted resilience. Thus, we can see that OVH can be a very non-resilient AS for certain sources AS, while being resilient for some others (e.g., OVH has origin-source resiliences $>= 0.8$ for roughly 20\% of the source ASes.) 

Similarly, Figure \ref{fig:topclient_hijack} shows the results of top 95 Tor Client ASes as source ASes, and the weighted curve only considers Tor Guard ASes. Again, we see greater heterogeneity in origin-source resiliences for both ASes than in the averaged case. Interestingly, AS12876 (ONLINE S.A.S) have lower resiliences than AS16276 (OVH) across all clients. This is also consistent with the results in Figure \ref{fig:hijack_as}, in which OVH has resilience 0.49 compared to ONLINE S.A.S's 0.39.

\subsection{Interception Resilience Results}
\label{intercept_results}


{\bf Using Tier-1 ASes as intercepting ASes.} Tier-1 ASes play an important role in Internet routing. They sit at the top level of the Internet hierarchy and carry a large amount of network traffic. Tier-1 ASes do not have any providers and are all fully peered with each other. Recall from Section \ref{interception_methodology} that in order to successfully intercept traffic of a true origin AS, the false origin AS needs to satisfy a \emph{safety condition} - it cannot announce the invalid route to its providers when its existing route to true origin AS is through a provider route. This condition puts Tier-1 ASes at a powerful position - Tier-1 ASes do not have any providers and thus can always announce the invalid route to all its neighbors (peers/customers), who will further propagate the announcement down to other ASes in the Internet hierarchy. On the contrary, ASes that are towards the bottom of the Internet hierarchy do not have much interception power. They have limited number of peers/customers, and most of their outgoing routes are through providers. Therefore, due to the difference in interception power, we only focus on measuring interception resilience to Tier-1 ASes as the attacking AS here instead of \emph{all} ASes as the attacking AS. 

\begin{figure}[ht!]
\centering
\includegraphics[width=.5\textwidth]{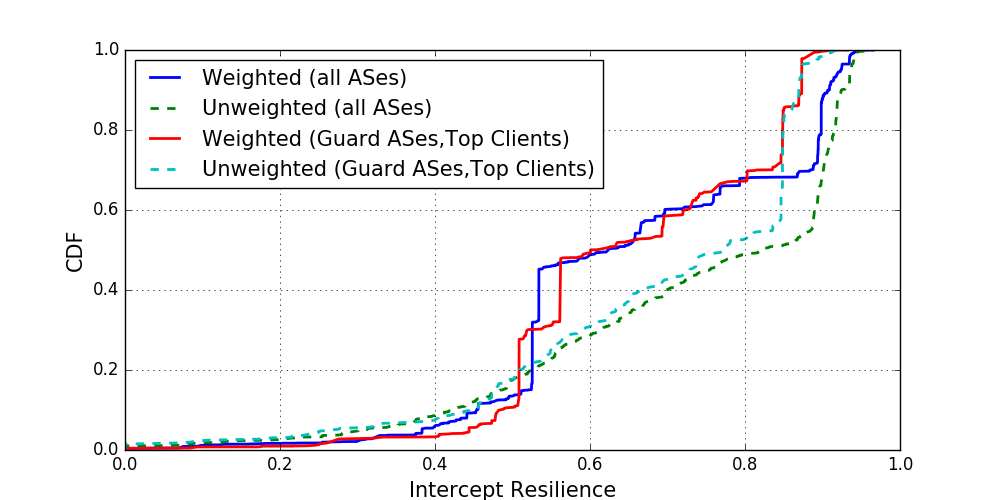}
\caption{Interception Resilience for Tor-related ASes}
\label{fig:intercept_as}
\end{figure}

\begin{figure*}[ht!]
\def\arraystretch{1.1}
\centering\captionsetup{aboveskip=6pt}
\begin{subfigure}{.5\textwidth}
  \centering\captionsetup{width=.9\linewidth}%
  \includegraphics[width=\linewidth]{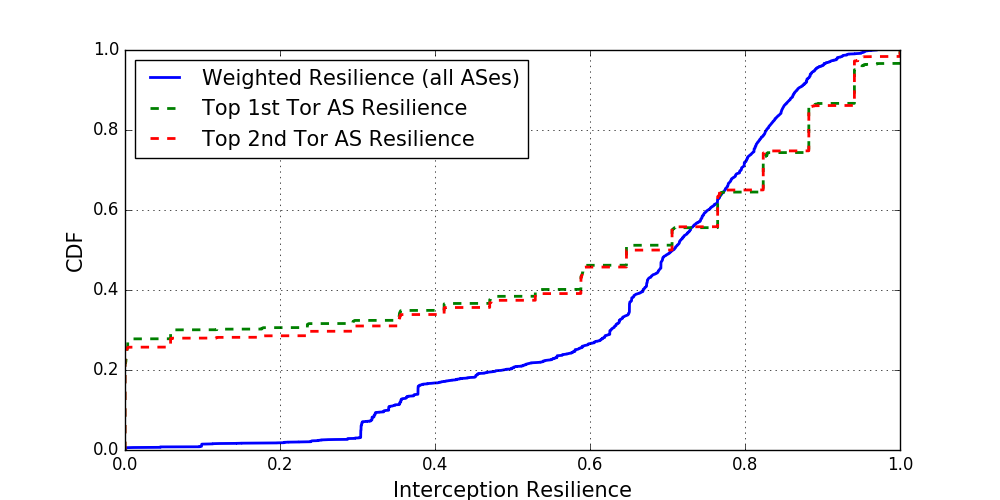}
  \caption{Interception Resilience from All Source ASes}
  \label{fig:all_intercept}
\end{subfigure}%
\begin{subfigure}{.5\textwidth}
  \centering\captionsetup{width=.9\linewidth}%
  \includegraphics[width=\linewidth]{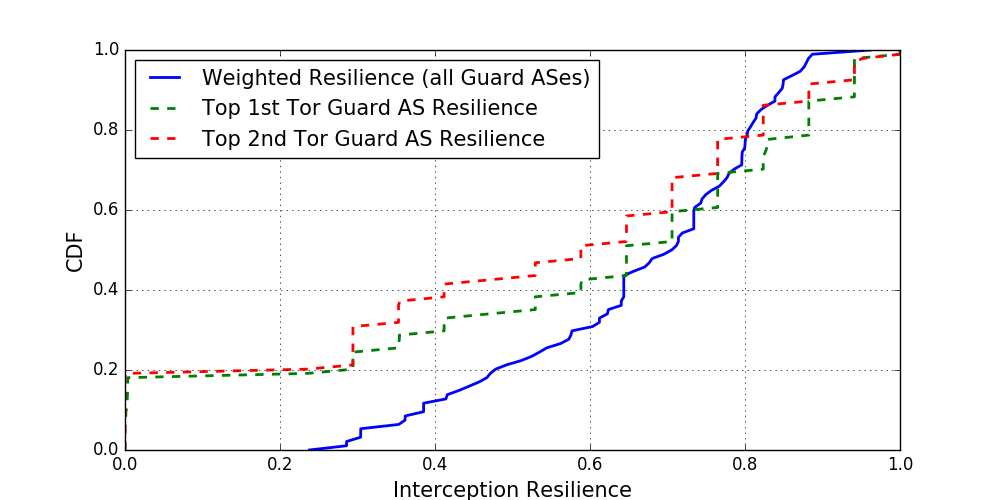}
  \caption{Interception Resilience from Top Client ASes}
  \label{fig:topclient_intercept}
\end{subfigure}
\caption{Interception Resilience for Top-Bandwidth ASes across Different Clients.}
\label{fig:intercept_client}
\end{figure*}

{\bf Interception resilience evaluation and results.} We used 17 Tier-1 ASes in our evaluation. \footnote{AS174, AS209, AS286, AS701, AS1239, AS1299, AS2828, AS2914, AS3257, AS3320, AS3356, AS5511, AS6453, AS6461, AS6762, AS7018, AS12956} Same as in the hijack resilience evaluation, we used the Tor consensus data and CAIDA AS topology data, both from January 2016. We first evaluate interception resilience of all Tor-related ASes, as shown in Figure ~\ref{fig:intercept_as}. We can see that the Tor ASes have higher resiliences to interception attacks than to hijack attacks, with roughly 50\% of the ASes having resilience higher than 0.8. The intuition behind this is that, even though Tier-1 ASes are at a position to intercept traffic, they also have longer paths from ASes that are close to the bottom of the hierarchy, which may prefer closer ASes with shorter paths instead of taking the longer paths to reach the Tier-1 ASes. Note that the resilience values are slightly lower for the Guard ASes (from top Client ASes) than all Tor ASes (from all source ASes). 

The \emph{weighted} distributions, though, show obviously lower resiliences, with the 50\% point at resilience value 0.6. This is, again, due to some high bandwidth Tor ASes which have low resiliences, e.g., AS16276 (OVH) only has resilience of 0.56. The analysis in Section \ref{hijack_results} on OVH's topological features in the AS hierarchy can also be applied here to explain its low resilience. 

We also plot the origin-source resilience distributions of each source/client AS for the Top 2 high bandwidth ASes (AS16276 and AS12876), as shown in Figure \ref{fig:intercept_client}. Consistent with our observations above, the two high bandwidth ASes have clearly lower origin-source resiliences across most of the source ASes, which attribute to their low origin resiliences. Although, as discussed in Section \ref{hijack_results}, they may have low resiliences for certain source ASes while having high resiliences for others, as reflected in the figure as well.

{\bf Consider resilience when choosing relay.} Tor clients choose relays based on relay bandwidth, and thus high-bandwidth relays have high chances of being chosen by Tor users. However, we have shown in this section that equally-specific attacks are real threats which have already affected the Tor network in the past, and the current bandwidth-based relay selection may choose very low-resilience relays for the clients. Therefore, guard relay selection that solely relies on relay bandwidth can expose many Tor users to the high risk of being compromised by active BGP attacks. This vulnerability motivates our work on incorporating AS resilience into guard relay selection, which we will delve into in Section \ref{sec:relayselection}. 


\section{Proactive Defense: \\Tor Guard Relay Selection}
\label{sec:relayselection}

Guard relays are at an important position in the Tor circuit, since they have direct connections with Tor clients. Strategic adversaries can discover the users using specific guard relays via BGP hijacks, and even perform traffic correlation analysis to deanonymize users via BGP interceptions \cite{sun2015raptor}. The attacks can be either more-specific prefix attacks or equally-specific prefix attacks. 
While more-specific prefix attacks affect the whole internet and mitigating such attacks could require cooperations from relay operators (e.g., moving relays into /24 prefix length), equally-specific prefix attacks affect a portion of internet, and Tor clients can possibly stay unaffected during such attacks by choosing the guard relay wisely and \emph{proactively} before any attack happens. In addition, as shown in Section ~\ref{subsec:equal_case_study}, equally-specific prefix attacks are real threats that have already affected Tor users in the past. To this end, we propose a new Tor guard relay selection algorithm that incorporates AS resilience to minimize the probability that Tor clients would be affected when their guard relays are under equally-specific prefix attacks. 

The following are the design goals of our guard relay selection algorithm. 
\begin{enumerate}
\item \emph{Mitigate equally-specific prefix attacks on Tor.} This is the main goal of the selection algorithm. The algorithm computes the AS resilience against prefix hijacks of all Tor guard relays from the client source AS, and prefers the ones that have higher resilience to minimize the likelihood that the client would be affected by a prefix hijack on its guard relay. 
\item \emph{Protect the anonymity of Tor clients.} In addition to lowering the possibilities of being hijacked, the algorithm should also protect the anonymity of Tor users by balancing preferences among relays and providing rigorously assessed anonymity bounds. 
\item \emph{Performance and load balancing.} The algorithm should incorporate relay bandwidth into the selection decision and avoid causing excessive traffic congestion on low bandwidth relays.
\end{enumerate}

\subsection{Guard Relay Selection Algorithm}
\label{subsec:selectionalgorithm}
We describe our Tor guard relay selection algorithm in detail with regards to two aspects: 1) choosing resilience metric and 2) incorporating resilience into relay selection.

\subsubsection{Choosing Resilience Metric}
\label{subsec:choose_resilience}
In general, ASes are more resilient to interception attacks than to hijack attacks, as shown in Section \ref{hijack_interception_measurement}. The interception resilience is a conservative measure of the probability of intercepting packets without additional set ups (e.g., VPN tunnels that send packets to the origin, or tunneling to a colluding AS which can then forward packets to the origin, etc.). However, there could be many cases where the probability of intercepting is higher. In other words, the interception resilience provides a \emph{upper bound} on resilience to packet interception. On the contrary, the hijack resilience considers a basic property that \emph{any} packet interception needs to satisfy, and hence provides a \emph{lower bound} on resilience to packet interception. For this reason, we will choose the hijack resilience to incorporate into the guard relay selection. 

\subsubsection{Incorporating resilience}
\label{subsec:as_resilience}
We defined the \emph{origin-source resilience} in Section \ref{hijack_methodology}, which represents the probability of a given source AS being resilient to attacks on a true origin AS. Here, the source AS is the AS where Tor client is located, and the true origin ASes are the ones which contain eligible Tor guard relays. Algorithm \ref{algo:calcres} describes how to calculate the origin-source resilience $R(i)$ of each Tor-related AS $i$ from the client AS $v$. 

Tor relay selection is bandwidth-aware and prefers high bandwidth relays. The probability of each relay $i$ being chosen is based on its default bandwidth $B(i)$. We offer a tunable parameter $\alpha$ in the relay selection algorithm, combining hijack resilience $R(i)$ and the default bandwidth $B(i)$. Each relay $i$ will be assigned a weight as following:
\begin{equation*}
W(i) = \alpha \times R(i) + (1 - \alpha) \times \bar{B(i)}
\end{equation*}
Note that, $B(i)$ is normalized to $\bar{B(i)}$, which is in $[0,1]$. when $\alpha$ is set to $0$, the relay selection becomes the same as bandwidth-only selection; while when $\alpha$ is set to $1$, the selection becomes resilience-only selection. 

\subsubsection{Randomization is needed}
\label{subsec:relay_randomization}
If we simply select the set of guard relays based on the probability of $W(i)/\sum W(i)$, an adversary can potentially run a relay that has an AS-level path with high local preferences and/or short path length to the Tor client, such that it has high resilience from the client AS as the source. Via this attack, an adversary obtains a high probability of being chosen. Furthermore, the Tor client might also be susceptible to fingerprinting attacks due to the differences in relay selection probabilities based on the AS-location of the client. An adversary that can observe the client for a long enough time may be able to infer the AS-location of the client based on its observed relay selection choices. Thus, we need to take into account these potential vulnerabilities and protect the anonymity of clients. Recall that the weight of a Tor relay depends on two components: (1) the resilience of the AS in which the relay is located, and (2) the relay's bandwidth. The relay's bandwidth is not specific to client locations, and thus would not reveal any client identities; in addition, due to resource constraints, it is not trivial to run a relay with significantly higher bandwidth than all other relays to obtain high probability of being chosen. On the other hand, AS resilience of relays \emph{is} client-specific, and requires much less resource to run a malicious relay with high AS resilience.

Instead of using resilience $R(i)$ for relay $i$ directly in the weight calculation, we first adjust it to $R(i)^{\prime}$ by calculating the estimated inclusion probability of the relay in a random sampling of size $(g \cdot N)$ using the algorithm proposed by Tille \cite{tille1996elimination}. Here, $N$ corresponds to the total number of Tor guard relays, and $g$ is a configurable parameter indicating the percentage of random sampling we want to perform. \emph{The intuition behind using Tille's algorithm is that we want to first pick $(g \cdot N)$ relays based on $R(i)$, and then randomly pick one among the selected relays.} Tille's algorithm provides an estimation of the probability of a relay being chosen given its $R(i)$.  The steps are as following:
\begin{enumerate}
\item For each relay $i$, $R(i)^{\prime} = \frac {k \cdot R(i)} {\sum_{j \in S} R(j)}$ in which $k$ is initially equal to the sample size $(g \cdot N)$ and set S initially includes all available relays. 
\item For each relay $i$, if $R(i)^{\prime} > 1$, $R(i)^{\prime} = 1$ and $k = k - 1$, and exclude relay $i$ from set S. 
\item Repeat the above process until each $R(i)^{\prime}$ is in $[0,1]$.
\item For each relay $i$, $R(i)^{\prime} = \frac{R(i)^{\prime}} {g \cdot N}$
\end{enumerate}

Note that when $g$ is set to $\frac 1 N$, then no random sampling will be performed, while if $g$ is set to $1$, then all relays will have the same $R(i)^{\prime}$ in their weights.

\subsection{Implementation on Tor}
\label{subsec:implementation}
Mapping the IP addresses of the Tor client and the Tor relays to their respective AS is necessary before we can compute AS resilience. In order to preserve the anonymity of the Tor client and not reveal its location to outside servers or anyone who can observe its communications, the client will perform the IP to ASN mapping locally by utilizing the Maxmind ASN database ~\cite{maxmind}, which can be included in the Tor download package. Note that the Maxmind GeoIP database for IP to Country mapping is already included in the Tor package and being used by the vanilla Tor client. In addition, the client will use the AS topology database from CAIDA \cite{caida} ($<$ 700KB compressed) for AS-level path inference in the resilience calculation. Here, we assume that  Maxmind GeoIP database and CAIDA AS topology database are both reliable sources. Note that, CAIDA only updates the AS topology database monthly, so the overhead incurred for downloading the most recent file is low. The detailed steps are as following:


\begin{enumerate}
\item If the Maxmind ASN file and AS topology file have not been downloaded, the Tor client will download the two files from Maxmind and CAIDA, respectively, and save them in the local data directory. Otherwise, the Tor client will check if the local AS topology file is up to date (updated monthly), and if not, then download the latest version.
\item The Tor client will perform IP to ASN mapping, and compute the AS resilience $R(i)$ of all candidate guard relays from the client AS as the source AS. 
\item The Tor client will perform random sampling on all candidate guard relays and adjust the resilience value to $R(i)^{\prime}$.
\item The Tor client will compute a weight for each candidate relay using formula $W(i) = \alpha \times R(i) ^{\prime} + (1 - \alpha) \times \bar{B(i)}$. 
\item The Tor client will proceed with the path selection. The remaining part of the circuit construction process stays the same as it is in Tor. 
\end{enumerate}

\subsection{Security and Anonymity Evaluation}
\label{subsec:security_eval}
We evaluate the security and anonymity of the Tor guard relay selection algorithm from three perspectives: (1) increasing the probability of a Tor client being resilient (unaffected) to a hijack attack on the Tor guard relay, (2) vulnerability to client fingerprinting attacks, and (3) rigorously assessing anonymity bound for a given Tor client using MATor ~\cite{backes2014nothing}.

\subsubsection{Probability of a Tor client being resilient to a hijack attack on Tor guard relay}
\label{subsec:attack_probability}
This is the main goal of the new relay selection algorithm. Let $P_{pick}(i)$ denote the probability that a Tor client will choose relay $i$ using our algorithm, and $P_{resilient}(i)$ denote the probability that a Tor client will stay unaffected if relay $i$ is being hijacked. $P_{resilient}(i)$ is essentially the same as the origin-source resilience described in Section ~\ref{hijack_methodology}. The aggregated probability of a Tor client being resilient to a hijack attack on guard relay can then be expressed as:
\begin{equation*}
\sum_{i \in \{\mbox{all guard relays}\}} P_{pick}(i) * P_{resilient}(i)
\end{equation*}

\begin{figure}[ht!]
\centering
\includegraphics[width=.5\textwidth]{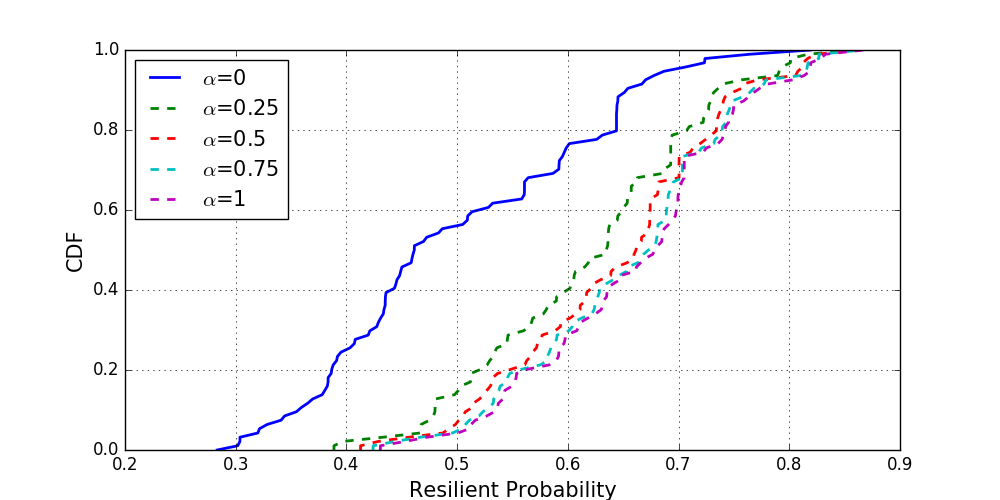}
\caption{Probability of being resilient to attacks with different $\alpha$ values}
\label{fig:resil_prob}
\end{figure}

We evaluate the probability for five values of $\alpha=\{0, 0.25, 0.5, 0.75, 1\}$, using $95$ top Tor client ASes \cite{johnson2013users} as the source ASes and Tor consensus data from January 2016. Figure \ref{fig:resil_prob} shows the result. Naturally, $\alpha=1$ has the highest probability of being resilient by an attack, with an average of 36\% increase. Note that, the algorithm benefits certain clients more than the others. For instance, if a client already has high probability of being resilient under the current relay selection algorithm, then its space for improvement would be low, as shown in Figure \ref{fig:resil_prob}. 

We then evaluate it with random sampling of $g=10\%$. We choose $10\%$ as the random factor here based on empirical evaluations of different $g$ values, and we found that the marginal benefit of a larger $g$ value does not compensate the loss in resilience to hijack attacks and performance. 

Table \ref{tbl_attack_reduction} shows the average relative percentage of improvement in the probability of being resilient to a hijack attack compared to $\alpha=0$, for both with and without random sampling. We can see that, even though the highest average is 36\%, the maximum percentage can be up to 166\% for certain clients. Also, there is only a slight decrease in percentage of improvement for higher $\alpha$ values with the 10\% random sampling. 

%
%
\begin{table}[ht!]
\begin{center}
\small
 \begin{tabular}{ p{5mm} | p{1.6cm} | p{1.6cm} | p{1.6cm} | p{1.6cm}}
    \hline
    $\alpha$ & Average (No sampling) & Average (10\% sampling) &  Maximum (No sampling) & Maximum (10\% sampling) \\ \hline
    0.25 & 27\% & 22\% & 114\% & 93\%\\
    0.5 & 32\% & 30\% & 144\% & 131\%\\
    0.75 & 35\% & 34\% & 158\% & 153\%\\
    1 & 36\% & 36\% & 166\% & 166\%\\
    \hline
  \end{tabular}
\end{center}
\caption{Percentage of improvement in resilience compared to $\alpha=0$ \label{tbl_attack_reduction}}
\end{table}

\subsubsection{Vulnerability to client fingerprinting attacks}
\label{subsubec:fingerprint}
There is a potential security tradeoff in the relay selection algorithm between vulnerability to prefix hijack attacks and vulnerability to fingerprinting attacks. We briefly discussed in Section ~\ref{subsec:relay_randomization} about fingerprinting a client location based on its preferences of relays in the long term. 
For instance, for a given relay, if client $a$ has 70\% probability of choosing the relay while client $b$ only has 30\% probability, then an adversary can observe the client's choice of relays over time to infer client information. The resilience component of our relay selection algorithm may be subject to such fingerprinting attacks, which we will evaluate here. 

We used an entropy-based anonymity metric, Shannon Entropy \cite{coifman1992entropy}, to evaluate the information leak. This metric considers the distribution of potential Tor clients of the connection (as computed by the attacker) and computes its Shannon entropy as:

\begin{equation*}
H_{Shannon} (I) = - \sum_i p_i log_2 p_i
\end{equation*}

where $p_i$ is the probability that for the given relay, client $i$ is the initiator of the connection. We consider the top 95 Tor client ASes \cite{johnson2013users} as potential clients, and focus on guard relays. Note that, the maximum entropy that can be achieved will be $\log_2 95 = 6.57$. 


Table \ref{tbl_client_entropy} shows the result. Bandwidth-only selection in Vanilla Tor ($\alpha = 0$) has maximum entropy $6.57$ for all relays, since the probability of it being chosen is the same across all clients, and thus does not leak any client information. With resilience-based selection, the entropy becomes lower. However, the loss in entropy is not significant - with $2.4\%$ loss when $\alpha = 0.25$, and $4.1\%$ loss when $\alpha = 1$. With the 10\% random sampling, this loss is further reduced down to $1.7\%$ loss when $\alpha = 0.25$, and $3.9\%$ when $\alpha = 1$. Furthermore, since Tor clients only select guard relays at bootstrapping time and would then use the same guard relays over several months (or until the relays become unavailable), so precise fingerprinting could not be done in a reasonably short time (without launching massive DoS attacks which cause guard relays to become unavailable). Especially given the very minor loss in entropy as shown in Table \ref{tbl_client_entropy}, the attacker will not be able to efficiently locate client ASes. Additionally, there could be hundreds of thousands or millions of clients in an AS, so even knowing the client AS still does not imply precise client information.

\begin{table}[ht!]
\begin{center}
\small
 \begin{tabular}{ p{5mm} | p{1.6cm} | p{1.6cm} | p{1.6cm} | p{1.6cm}}
    \hline
    $\alpha$ & Entropy (No sampling) & Percentage Reduction  & Entropy (10\% Sampling) & Percentage Reduction\\ \hline
    0 & 6.57 & - & - & -\\
    0.25 & 6.41  & 2.4\% & 6.46 & 1.7\%\\
    0.5 & 6.36  & 3.2\% & 6.39 & 2.7\%\\
    0.75 & 6.32  & 3.8\% & 6.35 & 3.3\%\\
    1 & 6.3 & 4.1\% & 6.31 & 3.9\%\\
    \hline
  \end{tabular}
\end{center}
\caption{Average Shannon Entropy. Note that when $\alpha=0$, entropy reaches maximum value of 6.57, indicating a completely uniform distribution.}
\label{tbl_client_entropy}
\end{table}


{\bf Recommended setting of $\alpha$ value.} 
We can see from Figure \ref{fig:resil_prob} that $\alpha=\{0.75,1\}$ does not have a significant advantage in increasing resilience over $\alpha=\{0.5\}$ in Section \ref{subsec:attack_probability}. Thus, we recommend using $\alpha=0.5$ as default, since it provides an obviously greater increase in resilience to attacks than $\alpha=0.25$, 
while the marginal benefit decreases as $\alpha$ continues to increase. $\alpha=0.5$ also offers a relatively higher entropy (and thus lower vulnerability to fingerprinting attacks) compared to bigger $\alpha$ values, as shown in Table \ref{tbl_client_entropy}. However, if the Tor client has some special configurations (e.g., clean up cached circuits and connect to new guards frequently instead of using the default guard relay configuration), then she should consider using a lower $\alpha$ value. In Section \ref{subsec:performance_eval}, we will show that $\alpha=0.5$ provides good performance as well. 

\subsubsection{Anonymity assessment}
Finally, we evaluate the anonymity of a given Tor client using MATor, a framework for assessing the degree of anonymity in Tor with rigorously proved anonymity bounds \cite{backes2014nothing}. Note that the anonymity notion in MATor is different from that in Section \ref{subsubec:fingerprint}: MATor considers a \emph{given client} and measures anonymity with \emph{different relays} that may be chosen by this given client, while the latter considers a \emph{given relay} and measures the entropy in the probabilities of it being chosen by \emph{different clients}.

\begin{figure}[ht!]
\centering
\includegraphics[width=.4\textwidth]{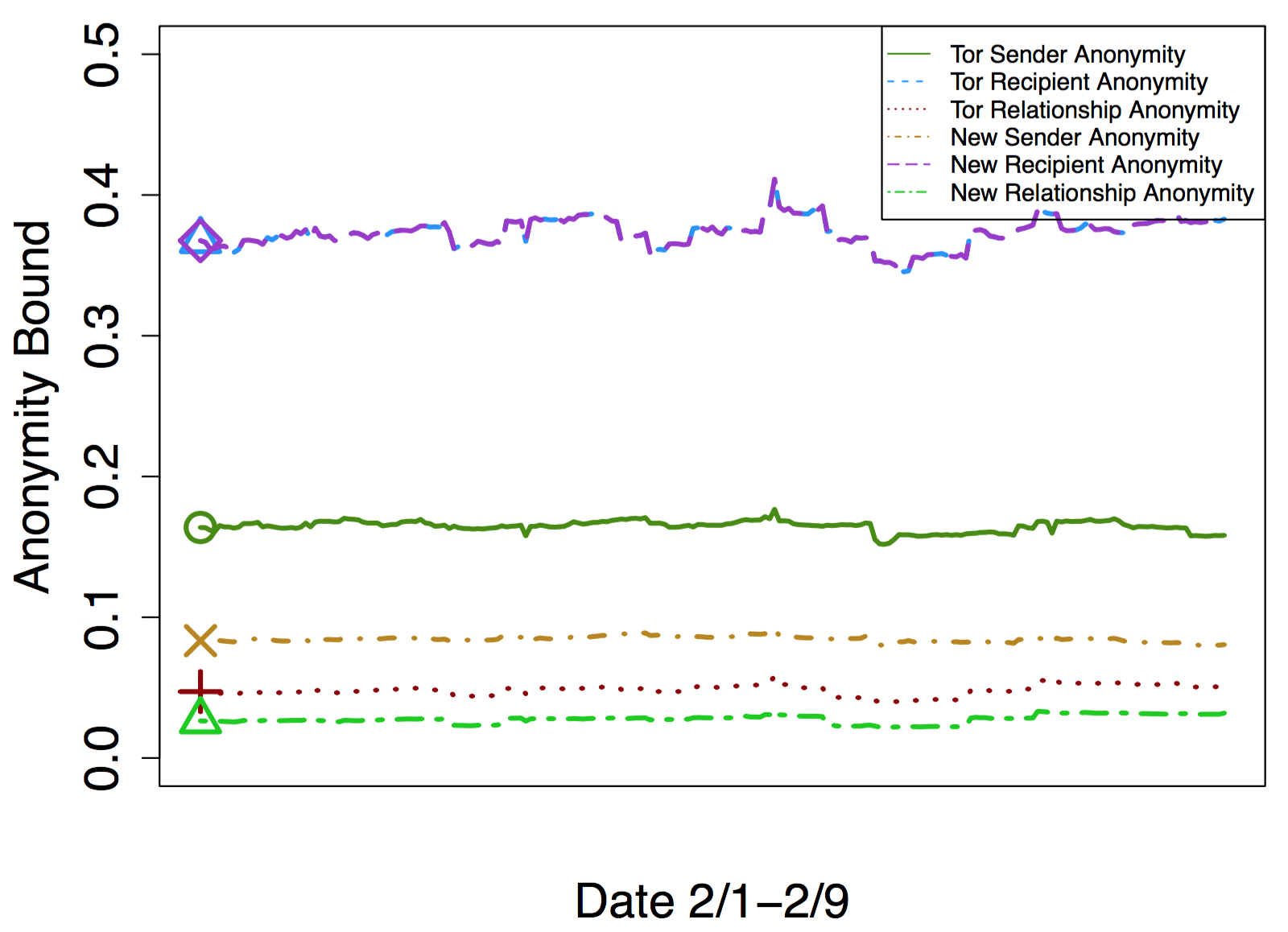}
\caption{MATor Anonymity Bound 2/1/2016 - 2/9/2016\label{fig_mator}}
\end{figure}

We implemented and integrated our new guard relay selection algorithm into MATor, and evaluated it in comparison with vanilla Tor. Note that we picked the top Tor client location AS6128 \cite{juen2012protecting} to evaluate here. We used MATor's default configuration of multiplicative factor $\epsilon=1.3$, ports setting of HTTPS+IRC vs. HTTPS, and $0.5\%$ of total nodes as compromised nodes (considering a worst-case adversary with a limit on the number of nodes it can compromise). We evaluated using Tor consensus files from 2/1/2016 - 2/9/2016 and server descriptor from February 2016. Figure ~\ref{fig_mator} shows the result. 

MATor evaluates three anonymity notions (sender, recipient, and relationship anonymity). The full details of the anonymity definitions are described in ~\cite{backes2014nothing}. \emph{The result shows that our new guard relay selection algorithm has tighter anonymity bounds on sender and relationship anonymities compared to current Tor path selection, indicating better anonymity guarantees.} The recipient anonymity remains the same as vanilla Tor, which is expected since we do not alter selection algorithm for exit relays. The intuition behind the better anonymity provided by our algorithm is that we redistribute the preferences for guard relays by factoring in relay resiliences. This avoids placing high trust in a small set of high-bandwidth nodes, and thus results in better anonymity.


\subsection{Performance Evaluation}
\label{subsec:performance_eval}

We implemented our new Tor guard relay selection algorithm by modifying Tor's source code, and evaluate on the Shadow emulator \cite{jansen2011shadow} for large scale and whole system network performance. We configured the Tor network in our simulation as in Table \ref{tbl_shadow_config}. Note that, this is the default configuration that comes with the Shadow Tor plug-in, which has been fine tuned by Shadow developers to model the Tor network.

\begin{table}[ht!]
\begin{center}
\small
 \begin{tabular}{ p{4cm} | p{2cm} }
    \hline
    Type & Number  \\ \hline
    Web Client & 360 \\
    Bulk Client & 40 \\
    Web Server & 100 \\
    Guard Relay & 14 \\
    Exit Relay & 10 \\
    Guard/Exit Relay & 5\\
    Middle Relay & 66\\
    \hline
  \end{tabular}
\end{center}
\caption{Shadow Configuration}
\label{tbl_shadow_config}
\end{table}


\begin{figure*}[ht!]
\def\arraystretch{1.1}
\centering\captionsetup{aboveskip=6pt}
\begin{subfigure}{.5\textwidth}
  \centering\captionsetup{width=.9\linewidth}%
  \includegraphics[width=\linewidth]{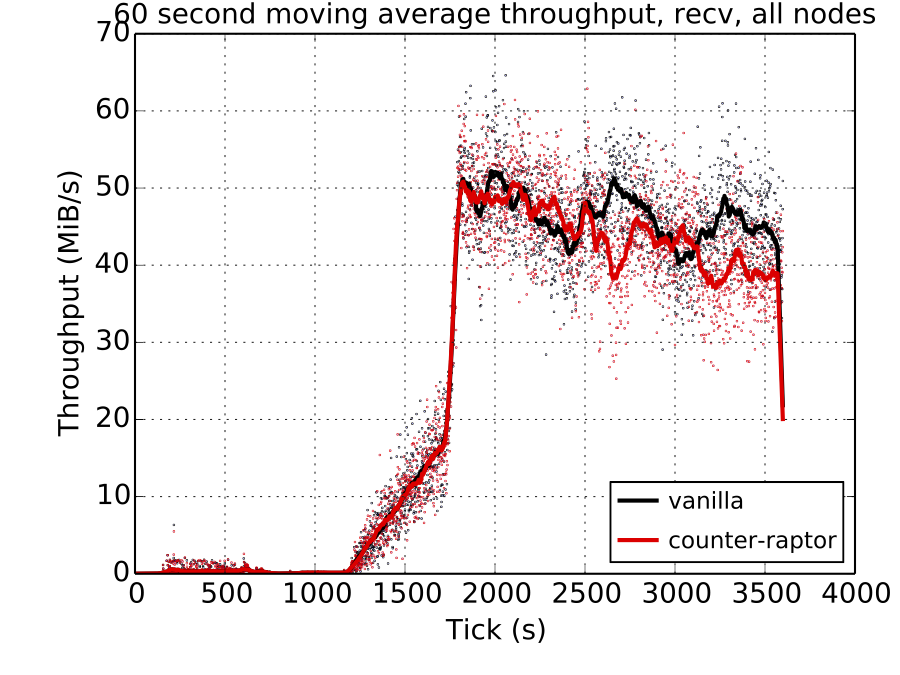}
  \caption{60 second average receiving throughput for all nodes.}
  \label{fig:throughput_recv}
\end{subfigure}%
\begin{subfigure}{.5\textwidth}
  \centering\captionsetup{width=.9\linewidth}%
  \includegraphics[width=\linewidth]{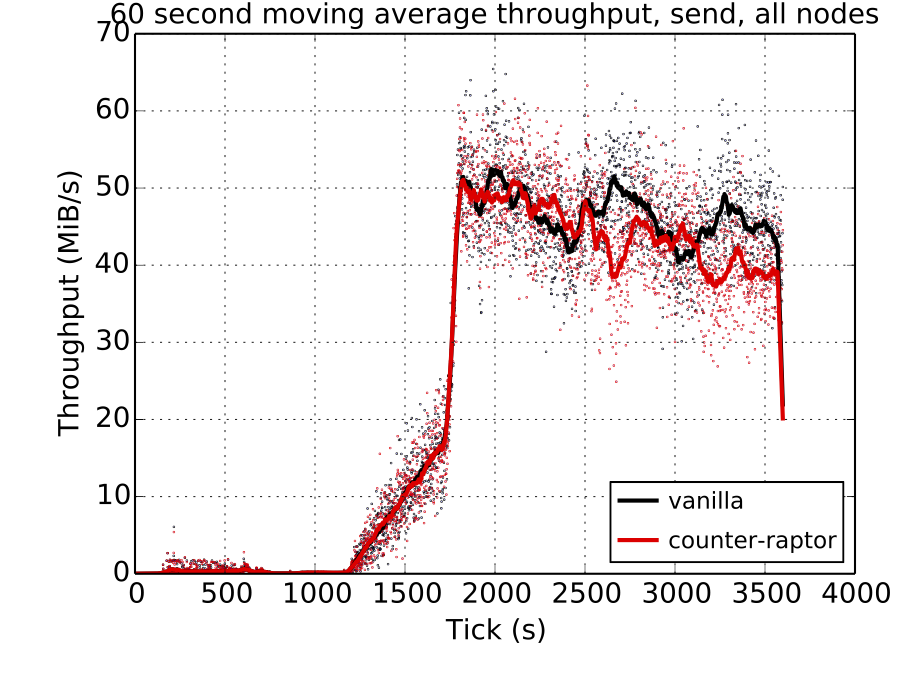}
  \caption{60 second average sending throughput for all nodes.}
  \label{fig:throughput_send}
\end{subfigure}
\label{fig:shadow}
\begin{subfigure}{.5\textwidth}
  \centering\captionsetup{width=.9\linewidth}%
  \includegraphics[width=\linewidth]{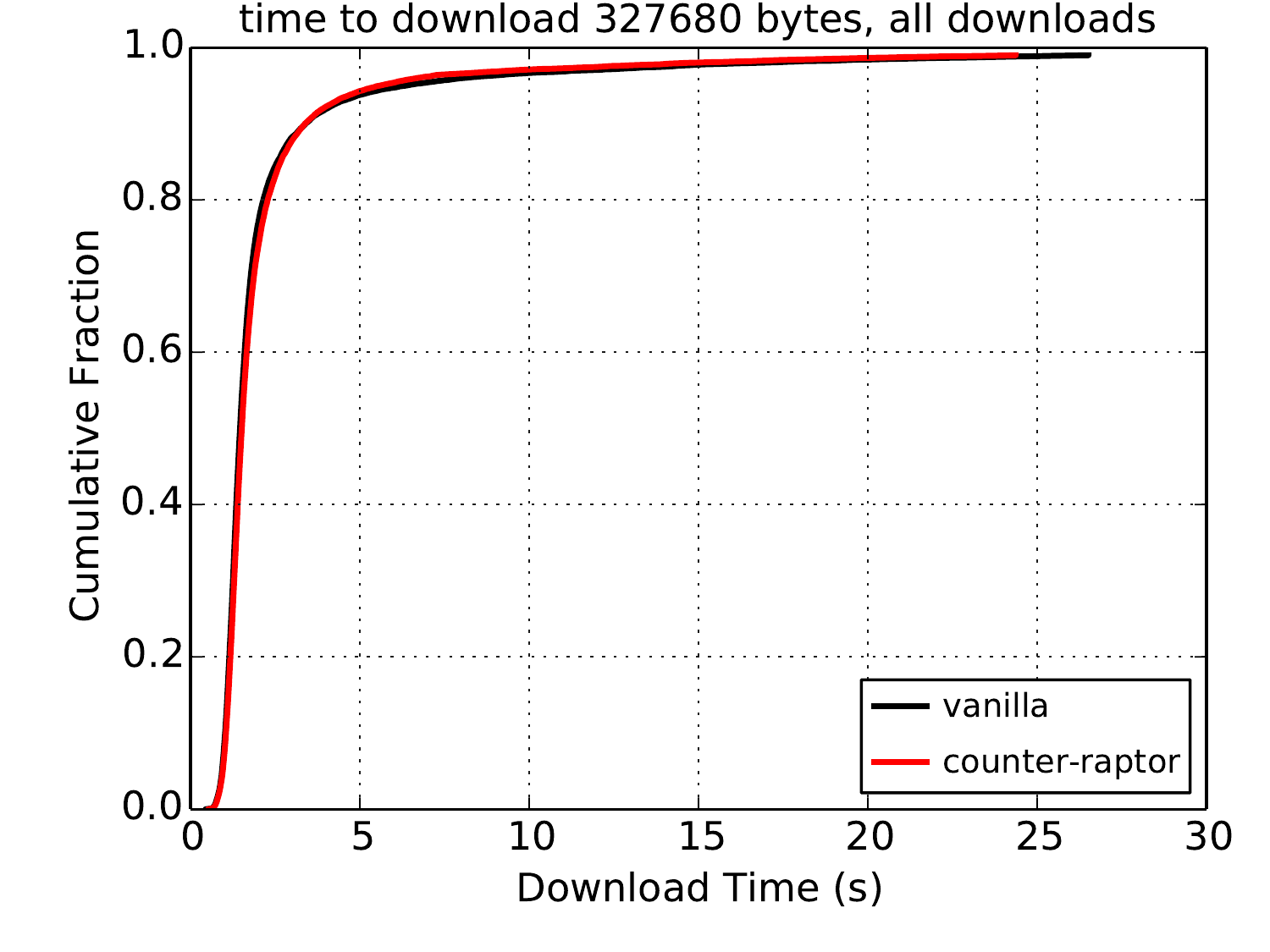}
  \caption{Download time for 320KB data.}
  \label{fig:all_download_kb}
\end{subfigure}%
\begin{subfigure}{.5\textwidth}
  \centering\captionsetup{width=.9\linewidth}%
  \includegraphics[width=\linewidth]{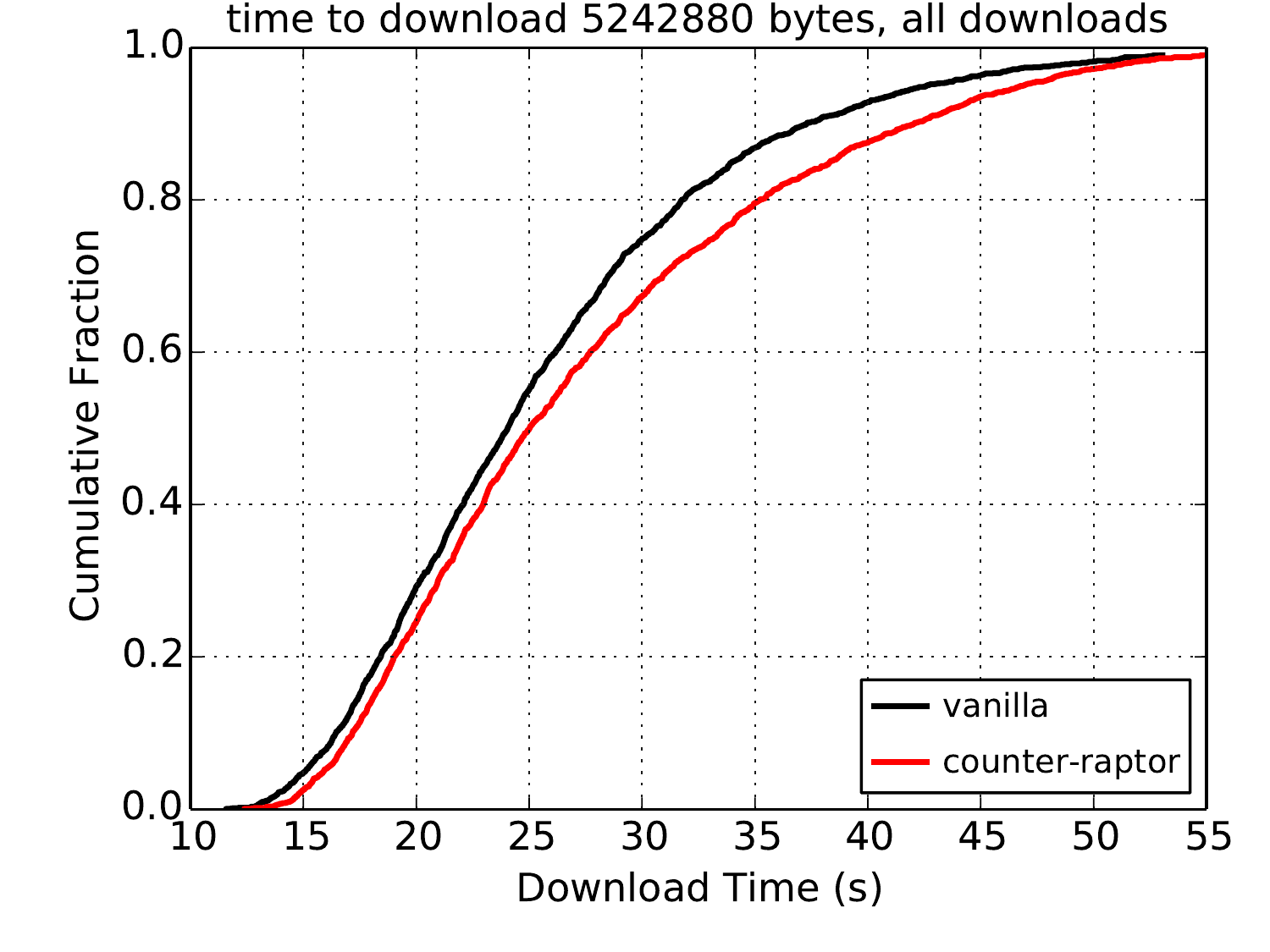}
  \caption{Download time for 5MB data.}
  \label{fig:all_download_mb}
\end{subfigure}
\caption{Large-scale evaluation of Tor guard selection algorithm on Shadow simulator.}
\label{fig:shadow}
\end{figure*}

Since our relay selection is location-dependent, we need to assign meaningful IP addresses to all the nodes. We used the IP addresses in the default Shadow configuration file for the relays, and we uniformly chose IP addresses from the 95 top Tor client locations \cite{johnson2013users} to the 400 Tor clients in our simulation. For simplicity, we only show the results for $\alpha=0.5$ here in comparison to Vanilla Tor. 

%
%

Figure \ref{fig:shadow} shows the network performance results from the emulation. Figure \ref{fig:throughput_recv} and Figure \ref{fig:throughput_send} shows the 60 second average receiver and sender throughput for all nodes, respectively. Counter-Raptor selection has almost the same throughput as Vanilla Tor during the simulation, while having two oscillations in the middle. Figure \ref{fig:all_download_kb} shows the download times of 320KB data. We can see that Counter-Raptor and Vanilla Tor have identical performance. Figure \ref{fig:all_download_mb} shows the download times of 5MB data. For this much larger data size, there is a minor increase in latency. 




As briefly explained in this Section, we do not restrict relay selection to a smaller set of relays and we give sufficient weight to the bandwidth ($\alpha=0.5$ here), which explain why our new guard selection algorithm only suffers minor performance loss compared to Vanilla Tor.

\section{Reactive Defense: \\BGP Monitoring System}
\label{sec:bgp}


The Tor guard relay selection algorithm in Section ~\ref{sec:relayselection} \emph{proactively} mitigates the affect of active BGP hijacks on the Tor client. In this section, we present a live BGP monitoring system that \emph{reactively} detects suspicious routing attacks that affect Tor relays. While there have been previously proposed BGP monitoring systems to detect prefix hijack attacks ~\cite{lad2006phas, zhang2008ispy, zheng2007light, qiu2007detecting, hu2007accurate, shi2012detecting}, our system is the first that has been tailored for the Tor network. We introduce a novel analytics-based approach for hijack detection, which is specifically designed and tuned for Tor. Our live monitoring system increases \emph{routing transparency} in the Tor network. Attackers will have to perform attacks in the public domain, as opposed to being stealthy.

Some relay operators have expressed interest in receiving alerts from our system upon routing anomaly detection on their relays.  This would be beneficial to relay operators as it would allow them to mitigate any attack much quicker than without the use of our system.  Therefore, fewer Tor users would be affected by the hijack because the attack would last a shorter amount of time.  Additionally, Tor users and others can also subscribe to our system to receive alerts. This could be helpful to Tor users by not only increasing the transparency in the Tor network, but also allowing them to select relays that are not being hijacked (in the case that they do not follow the default Tor settings, and select the relays they use).  Our system will be available to both relay operators and Tor users (or any subscriber) when we make our system publicly available.

\subsection{System Design}
\label{subsec:design}

\paragraph{Collecting Monitoring Data}
A BGP monitoring system on Tor requires information about current Tor relays.  The Tor Project releases up-to-date information about current running relays every hour. Our system automatically fetches this consensus data\footnote{\url{https://collector.torproject.org/recent/relay-descriptors/consensuses/}}. We focus on Tor guard relays and exit relays, which reside at the two ends of the communication path and can easily be the target of an adversary. Furthermore, since we focus on AS-level adversaries, it is unnecessary to monitor each individual relay by its IP address. Instead, we monitor the /24 prefixes which contain Tor guard and exit relays. There is no need to monitor a more specific prefix than /24, since generally /24 is the longest prefix accepted in a BGP announcement.

We pull a live stream of BGP announcements and withdrawals from BGPStream~\cite{bgpstream}, an open source framework for live BGP data.  We filter the BGP data to focus on the prefixes that contain a Tor guard or exit relay, as well as all the sub-prefixes up to the length /24 in order to detect sub-prefix (a.k.a more-specific prefix) hijack attacks. 

We use IP to ASN mappings from Team Cymru \cite{teamcymru} to obtain AS ownerships of the prefixes that contain Tor guard/exit relays. Some prefixes are owned by an organization with multiple AS numbers, so we take this into consideration and store all AS origins of these prefixes.  One caveat of using Team Cymru is the potential inaccuracy and incompleteness of the data; the system could also be augmented to check multiple registries and compare results.

\paragraph{Detecting Routing Anomalies}
We develop a framework for hijack detection, which utilizes two different types of techniques to check if any BGP activity involving Tor relays is anomalous, as following: 

\begin{enumerate}
\item Origin AS check. We compare the origin AS in the live BGP data against the owner AS in the Team Cymru registry {\it in real-time}. If there is a mismatch, we flag the BGP update and the prefix as suspicious. 
\item Analytics-based detection. We use two novel detection analytics based on the frequency and time features of BGP updates; because there is a significantly smaller number of updates that include a Tor relay in comparison to all BGP updates, this analytics-based approach is effective for Tor, and produces a more reasonable amount of false positives (as discussed in further sections). If a BGP update for a prefix falls under the tuned threshold of either analytics, we flag the BGP update and prefix as suspicious. 
\begin{enumerate}
\item Frequency Analytic.  Routing attacks can be 
characterized by an AS announcing a path once (or extremely rarely) to a prefix 
that it does not own. The frequency analytic detects attacks that exhibit this behavior. 
It measures the frequency of each AS that originates a given prefix; if the frequency is 
below some threshold, then it could be a potential hijack attack.
\item Time analytic.  Many real-world attacks 
last a relatively short amount of time in comparison to life span of a prefix~\cite{indiahijack, syriahijack,indosat2014, malaysialeak}. The time analytic measures the amount of time each 
path to a prefix is announced for; if the amount of time is below some threshold, 
then there is the possibility of it being a routing attack. 
\end{enumerate} 
We select the threshold values for the frequency and time analytics, respectively, by evaluating the system detection accuracy on historical BGP data with known attacks. 
\end{enumerate} 

\paragraph{Mitigating Hijack Attacks}
Once a suspicious update is flagged by a combination of techniques in the framework, the relay(s) that are contained in the flagged (potentially hijacked) prefix are blacklisted.  Relays that are blacklisted should not be used as guards for a period of time.  Our system can detect when a route to the victim prefix has returned to normal, and will remove the prefix from the blacklist at that point.  This provides some attack mitigation by preventing use of the guard when it could potentially be hijacked.  Additionally, this design does not force relays to be blacklisted forever, and relays can be used once deemed safe again.

\subsection{Deployment and Data Analysis}
\label{data_analysis}
The BGP monitoring system has been running since February 1st, 2016. In this section, we analyze the data for the whole month of February, and tune the threshold values for the frequency and time analytics.  The threshold is calculated based on all BGP updates for the month of February that include a prefix that contains at least one Tor relay.  We tune the threshold on an entire month worth of data because any potential attack in the data is a source of pollution; an attack is less likely to stand out as anomalous in a small amount of data as compared to a larger amount of data, such as a months worth of data.  In Sections \ref{sim} and \ref{real} we apply these tuned threshold values to the months of March, April and May of 2016. Therefore, data collected in February serves as training data, and data collected in March, April, and May serves as test data.

During the month of February, we assume there were no hijack attacks that affected the Tor network, and therefore assume there are no attacks in our monitored data from the month.  Our detection analytics are run in real-time on an hourly basis in addition to analyzing the results of the real-time origin AS check for two purposes: 1) threshold tuning, and 2) false positive analysis.  We present the data analysis results of each detection method according to the framework described in Section \ref{subsec:design}.

{\bf Origin AS Check.} This checking is done in real time as the live BGP stream comes into our system. We compare the origin AS from the BGP announcement in the live data stream with the owner AS of the prefix from Team Cymru registry. If there is a mismatch, then we log the prefix and the origin AS from the BGP announcement. Table \ref{tab:summary} shows the result. 

\begin{table*}[ht!]
\begin{center}
\center
\small
    \begin{tabular}{| l | p{2cm} | p{2cm} | p{2cm} | p{2cm} | p{2cm} | p{2cm} |}
    \hline
    Month & Total \# of Tor Updates & Total \# of Origin AS Conflict Updates & \# of Unique Prefixes in Conflicts & \# of Unique ASes in Conflicts & \# of Total Unique Prefixes & \# of Total Unique Origin ASes \\ \hline \hline
    February & 1401633 & 84148 & 139 & 71 & 2195 & 837 \\ \hline
    March & 1077098 & 49025 & 164 & 79 & 2357 & 858 \\ \hline
    April & 1691325 & 326586 & 369 & 88 & 2453 & 859 \\ \hline
    May & 2403680 & 22267 & 79 & 52 & 1954 & 783\\ \hline
    \end{tabular}
\end{center}
\caption{Summary statistics for conflicting origin AS updates.}
\label{tab:summary}
\end{table*}

From Table \ref{tab:summary}, we can see that the total number of BGP updates with mismatching origin AS is large in comparison to the number of unique prefixes in these mismatching cases, with an even smaller number of unique ASes involved in these prefixes. There are many duplicate BGP updates over time, which will all cause an alert in the system; future work includes implementing a known ``benign'' set of (prefix, origin AS) pairs, such that duplicates do not trigger an additional alert.  Additionally, if we know that a ``mismatch'' AS origin is indeed authorized to make the announcement for the prefix (e.g., we may get this information from relay operators), then we can add this exception to our ``benign'' list so that it would not trigger an alert again when it sees the same announcement. Furthermore, we may be more interested to know how many \emph{new} mismatches occur (rather than the frequently recurring ones), which may not be anomalous since routing attacks usually last for a short time and do not exhibit a repeated pattern over a long period of time. For example, there were 164 unique mismatching prefixes in March and 79 unique ASes involved with the prefixes, but many of these mismatches also appeared in February. If we calculate the number of \emph{new} mismatching prefixes that appeared in March (but not in February), the number goes down to 55 unique prefixes and 25 unique ASes. Additionally, while more than a hundred prefixes may appear to be a large number of alerts, it is important to remember that it is the total number of alerts for an entire month, which averages to just a few alerts per day. To further reduce the number of false positives, the origin AS check can be combined with the new analytics. 

{\bf Frequency Analytic.} The frequency analytic is calculated in real-time, automatically once per hour; this time increment can be reduced from hourly to a per-minute basis. For our data analysis, and to tune the threshold value, we applied this analytic to the data from the month of February, with varying threshold values.  The threshold value corresponds to the ratio of (total number of times a given prefix is announced by a given AS)/(total number of times a given prefix is announced by all ASes). As we expected, the number of false positives is directly related to the threshold value that is set for the analytic; the higher the threshold value, the more false positives are reported.  On the other hand, setting the threshold value too low can cause false negatives (actual attacks that are not detected).  

We varied the threshold value from 0.000 to 0.004.  The false positive rate remained at 0\% until the threshold value was raised to 0.003, at which point it became 0.05\%.  Therefore, we select a threshold value of 0.0025 for the frequency analytic when we apply the analytic to future (test) data.  

{\bf Time Analytic.} Similar to tuning the threshold value of the frequency analytic, we tuned the threshold of time analytic using the data collected throughout February.  We applied the time analytic while varying threshold values in order to determine the optimal threshold.

We varied the threshold from 0.00 to 0.08.  The false positive rate remains at 0\% until the threshold value is raised to 0.07, at which point it became 0.05\%.  Therefore, the threshold value for the time analytic is 0.065, which will be used in our evaluation on data from March, April, and May.  

\subsection{Evaluation: Simulated Attacks}
\label{sim}
After tuning threshold values for our detection analytics, we evaluate our monitoring system by testing these values on data collected by the system during the months of March, April, and May.  Our evaluation should measure: 1) how accurately the detection mechanisms can detect attacks, and 2) how usable is the system (in terms of false positive rate), given that it alerts a subscriber when attacks are detected.  To measure this, we analyzed past real-world hijack attacks, and modeled simulated attacks after them.  These simulated attacks were injected in the off-line data that our monitoring system recorded.  When deciding which prefix to hijack, we randomly selected a prefix already contained in the monitoring data, and we used the false origin associated with the real-world hijack.  All attacks are equally-specific hijack attacks, as more-specific prefix hijack attack detection has previously been studied \cite{ballani2007study, lad2006phas, zheng2007light, shi2012detecting}.  The following are brief descriptions of the real-world attacks after which the simulations are modeled.

\begin{enumerate}
\item Syrian Telecommunications Establishment (STE) hijack in 2014~\cite{syriahijack}. We injected 3 BGP updates into our data to make it appear that an attack occurred for four minutes on March 23, 2016. The hijacked prefix was 185.15.244.0/22, and it was an equal-length prefix hijack attack. 
\item Korean Weather Service hijacked US Climatic Center (climate.gov) in 2014~\cite{many_hijack}.  We injected 10 BGP updates into the March 30th-31st data for hijack that lasted 25 hours.  The hijacked prefix was 103.56.207.0/24.
\item Windstream hijacked a SaudiNet prefix in 2014~\cite{many_hijack}.  We injected 2 updates in the March 31st data to represent a 1 hour hijack of prefix 104.37.192.0/24.
\item Windstream hijacked a Hadara Gaza prefix in 2014~\cite{many_hijack}.  To model this attack, we hijacked prefix 195.254.135.0/24 on April 30th for 8 hours.  We injected an update once per hour for a total of 8 updates.
\item Windstream hijacked of Advania Iceland prefix in 2014~\cite{many_hijack}.  We injected 9 updates over the course of 9 hours on April 30th to hijack prefix 89.187.128.0/19.  The malicious BGP announcements had a shorter path than the true announcements.
\item Windstream hijacked two China Telecom prefixes in 2014~\cite{many_hijack}.  We modeled these two attacks separately on April 30th by injecting 9 updates over 9 hours for prefixes 77.245.144.0/20 and 151.100.0.0/16.  
\item INEA S.A. hijacked a US Department of Defense prefix in 2014~\cite{many_hijack}.  This attack was simulated by injecting 2 updates on April 30th for prefix 107.181.174.0/24 for a total time of 17 minutes.
\item A2B Internet hijacked a network associated with Bitcoin in 2015~\cite{mada_hijack}. We injected 4 updates on May 14th announcing prefix 193.200.241.0/24 for an attack that lasted 11 hours.
\end{enumerate}

\begin{table*}[ht!]
\begin{center}
\center
\small
    \begin{tabular}{| c | c | c | c | c | c | c |}
    \hline
    Attack & Month & Prefix & Hijacking ASN & True Origin ASN & \# of Updates Injected & Length of Time of Hijack  \\ \hline \hline
    1) & March & 185.15.244.0/22 & 29256 & 24961 & 3 & 4 minutes\\ \hline
    2) & March & 103.56.207.0/24 & 10063 & 58477 & 10 & 25 hours \\ \hline
    3) & March & 104.37.192.0/24 & 7029 & 36077 & 2 & 1 hour \\ \hline
    4) & April & 195.254.135.0/24 & 7029 & 38935 & 8 & 8 hours \\ \hline
    5) & April & 89.187.128.0/19 & 7029 & 35592 & 9 & 9 hours \\ \hline
    6) & April & 77.245.144.0/20 & 7029 & 42868 & 9 & 9 hours \\ \hline
    7) & April & 151.100.0.0/16 & 7029 & 137 & 9 & 9 hours \\ \hline
    8) & April & 107.181.174.0/24 & 13110 & 46562 & 2 & 17 minutes \\ \hline
    9) & May & 193.200.241.0/24 & 51088 & 51167 & 4 & 11 hours \\ \hline
    \end{tabular}
\end{center}
\caption{Summary statistics for the simulated attacks modeled after real-world attacks.}
\label{tab:attack_stats}
\end{table*}

Table \ref{tab:attack_stats} shows the summary statistics and characteristics for the attacks that were injected into the March, April, and May data. 

After simulating the real-world attacks, we apply our detection mechanisms hourly and evaluate them for false positives and false negatives.  We discuss the accuracy and coverage of each defense mechanism individually.

{\bf Origin AS Check.} The origin AS check successfully detects all of the injected attacks, since they all triggered the mismatching AS origin. Similar to the results presented in Section \ref{data_analysis}, this check produces a large number of false positives, but the false positive rate can be reduced by analyzing unique origin ASes or unique prefixes being announced.  The system can retain a list of known, non-malicious (prefix, origin AS) pairs, such that duplicates do not trigger an alert.  Additionally, this check can be combined with the frequency and time analytics for a more precise set of alerts.  If either the frequency or time analytic also flag the (prefix, origin AS) pair, then the user is alerted to the potentially malicious update.

{\bf Frequency analytic.} We applied the frequency analytic in real-time to each hour in March, April, and May, with a threshold value of .0025 (as determined in Section \ref{data_analysis}).  The frequency of AS A announcing prefix P was calculated based on the previous month of data for the hour being analyzed.  For example, the first hour of April was analyzed with respect to the frequencies in the month of March. 

This analytic was able to detect all attacks described in Table \ref{tab:attack_stats}, and therefore produced 0 false negatives.  Each attack was detected within the first hour of the first announcement; for attacks that lasted longer than a single hour, the attacks were still flagged in subsequent hours.  

A common issue with monitoring and detection systems is the amount of false positives produced.  If the false positive rate is too high, then the system is unusable.  The frequency analytic produced very few false positives; on average, over the 2,616 hours monitored, the false positive rate was .38\%.  We saw that 99.9\% of the hours monitored produced 0 false positives.  

The results for the frequency analytic highlight an important characteristic about the Tor network: most prefixes are announced by a single AS in all updates, causing the frequency of the (prefix, origin AS) pair to most commonly be 1.0. This suggests that this analytic is suitable for use in monitoring BGP activities on Tor.

{\bf Time Analytic.}  We applied the time analytic with the threshold value of .065 (determined in Section \ref{data_analysis}) to every hour of data in March, April, and May.  As with the frequency analytic, the timing information of announcements in the current were analyzed with respect to the previous month's timing information.  

In terms of false negatives, the time analytic detected all of the simulated attacks.  Similar to the frequency analytic, the false negative rate was 0.  This analytic exhibited a low percentage of false positives at .19\% on average across all hours monitored.  As with the frequency analytic, 99.9\% of the hours in the dataset had 0 false positives.  

Again, similar to the frequency analytic, the time analytic results in very low false positives, indicating that it is also well-suited for monitoring the Tor network.  Both of these analytics help reduce the false positives produced by the AS check, while still flagging the true positives.

\subsection{Evaluation: Real-World Attack}
\label{real}
In addition to the simulated attacks, we also performed a real-world BGP hijack attack on prefixes that we owned for the duration of our experiment. Note that the IP range and AS number we used to perform hijack attacks were temporarily subleased to us for our experiment. 
This was a separate experimental setup from the simulated attacks presented in the previous section.

We announced the prefix 184.164.226.0/23, which we temporarily owned, using the PEERING testbed. \cite{schlinker2014peering}. It allows us to establish BGP connectivity with other ASes by proxying our announcement via dozens of deployed sites in the world. In order to perform the hijack attacks, we used two PEERING sites: AS2637 (GATECH) and AS226 (ISI). We first announced the prefix using AS2637, and also added the prefix to the hourly list of prefixes that contain Tor relays, which is used to filter out BGP updates that contain Tor relays (as described in Section \ref{subsec:design}). On May 16th, 2016, we performed an equally-specific prefix hijack for about 5 minutes by announcing the prefix using AS226, which is modeled after a real-world hijack attack~\cite{indiahijack, syriahijack,indosat2014, malaysialeak}. During the hijack, the origin for the prefix in some of the BGP announcements became AS226. 

All three detection mechanisms -- origin AS check, frequency analytic, time analytic -- flag this as an attack during our real-time monitoring and the hourly analytics.  While our attack can detect all the simulated attacks, this shows that it can also detect real-world attacks.  

\subsection{Adversarial Knowledge of Detection Techniques}
In order to bypass our analytics detection, the adversary needs to make the false announcements both i) more frequent so it will be higher than the frequency analytic threshold, and ii) for longer time so it will be higher than the time analytic threshold. These two would be intrinsically hard to achieve at the same time - routing attacks are usually short, since the longer/more frequent the attack is, the much less stealthier it becomes. Especially given that we use the previous month's data as the threshold basis, it would be very hard to overcome. Furthermore, the system will flag the suspicious announcement the first time it appears, so if more such suspicious announcements arrive, they would not be used for the analytics to avoid polluting the data.  By actively omitting subsequent (redundant) attack data, this system can defend against such strategic poisoning attacks.



\section{Discussion}
\label{sec:discussion}

{\bf Accuracy of AS path inference.}  
Part of our AS resilience calculation involves AS-level path inference from the network topology. Recent work has shown that path inferences using local preference and shortest path may not be completely accurate \cite{juen2015defending}, and thus path selection algorithms \cite{nithyanand2016measuring} that rely on the accuracy of AS path inferences could be affected. However, we only use path inference as an indicator of network connectivity to calculate origin resilience instead of predicting and replying on any \emph{precise} routes. Thus, our resilience calculation is robust to a certain degree of AS path inference inaccuracy and/or AS path churn. 

{\bf Reliability of BGPStream.}
BGPStream aggregates all of the BGP prefix updates seen by RouteViews and RIPE collectors. If some set of the collectors are manipulated or corrupted, the other collectors can still be used and the monitoring system will still be effective. If the aggregator (BGPStream) is manipulated or corrupted, data can still be verified against the collectors through RouteViews or RIPE, and the original data can be restored and analyzed using the analytics in our system. 

{\bf Comparing and Combining the Detection Techniques.}
As we can see from Section \ref{sec:bgp}, the origin AS check is successful at catching the hijack attacks (true positives), while resulting in a significant number of false positives. We discussed two potential ways of eliminating false positives in Section \ref{data_analysis}, which include getting input from relay operators and marking certain prefix announcements as "benign", as well as focusing on \emph{new} mismatching cases rather than recurring ones. We can combine the origin AS check with our frequency and time analytics to achieve even higher accuracy. In contrast to the origin AS check, the analytics-based detection methods result in low false positive rates, which are used as an additional filter on the alerts triggered by the origin AS check to eliminate false positives.

\section{Conclusion}

In this work, we have presented proactive and reactive countermeasures to safeguard Tor against active BGP routing attacks.  First, we evaluated the Tor network for its current state of resilience to hijack and interception attacks.  We observed that some ASes with high Tor bandwidth have relatively low resilience. Next, we presented a new Tor guard relay selection algorithm that proactively mitigates routing attacks. The algorithm successfully increases the probability of a Tor client being resilient to prefix hijack attacks. Finally, we presented a live monitoring system that uses multiple new detection mechanisms to alert subscribers to potential hijack attacks happening in real-time.  We evaluated the monitoring system, and found that it was able to detect simulated attacks modeled after real-world attacks, as well as a real hijack attack (performed by us), with a negligible false positive rate.  Overall, our work is the first work on \emph{proactively} mitigating active routing attacks on Tor, and the first on presenting a real-time monitoring system tailored for Tor. \\

\textbf{Source Code:} The source code of Counter-RAPTOR guard relay selection algorithm is available at \url{https://github.com/inspire-group/Counter-Raptor-Tor-Client}.

\section*{Acknowledgment}
The authors would like to thank Ethan Katz-Bassett, Brandon Schlinker and Italo Cunha for support on the PEERING testbed. Thanks to Philipp Winter, Roger Dingledine, Matthew Wright, Moritz Bartl and Ryan Wails for helpful discussions. Special thanks to Aaron Johnson for providing invaluable feedback. This work was supported by the National Science Foundation under grants CNS-1423139, the Open Technology Fund through the Information Controls Fellowship, and the Department of Defense (DoD) through the National Defense Science \& Engineering Graduate Fellowship (NDSEG) Program.

\balance

\bibliographystyle{IEEEtran}
{\bibliography{sigproc}}

\begin{thebibliography}{10}
\providecommand{\url}[1]{#1}
\csname url@samestyle\endcsname
\providecommand{\newblock}{\relax}
\providecommand{\bibinfo}[2]{#2}
\providecommand{\BIBentrySTDinterwordspacing}{\spaceskip=0pt\relax}
\providecommand{\BIBentryALTinterwordstretchfactor}{4}
\providecommand{\BIBentryALTinterwordspacing}{\spaceskip=\fontdimen2\font plus
\BIBentryALTinterwordstretchfactor\fontdimen3\font minus
  \fontdimen4\font\relax}
\providecommand{\BIBforeignlanguage}[2]{{%
\expandafter\ifx\csname l@#1\endcsname\relax
\typeout{** WARNING: IEEEtran.bst: No hyphenation pattern has been}%
\typeout{** loaded for the language `#1'. Using the pattern for}%
\typeout{** the default language instead.}%
\else
\language=\csname l@#1\endcsname
\fi
#2}}
\providecommand{\BIBdecl}{\relax}
\BIBdecl

\bibitem{dingledine2004tor}
R.~Dingledine, N.~Mathewson, and P.~Syverson, ``Tor: the second-generation
  onion router,'' in \emph{Proceedings of the 13th conference on USENIX
  Security Symposium-Volume 13}.\hskip 1em plus 0.5em minus 0.4em\relax USENIX
  Association, 2004, pp. 21--21.

\bibitem{tormetrics}
``Tor metrics,'' \url{https://metrics.torproject.org/}.

\bibitem{shmatikov2006timing}
V.~Shmatikov and M.-H. Wang, ``Timing analysis in low-latency mix networks:
  Attacks and defenses,'' in \emph{Computer Security--ESORICS 2006}.\hskip 1em
  plus 0.5em minus 0.4em\relax Springer, 2006, pp. 18--33.

\bibitem{syverson2001towards}
P.~Syverson, G.~Tsudik, M.~Reed, and C.~Landwehr, ``Towards an analysis of
  onion routing security,'' in \emph{Designing Privacy Enhancing
  Technologies}.\hskip 1em plus 0.5em minus 0.4em\relax Springer, 2001, pp.
  96--114.

\bibitem{edman2009awareness}
M.~Edman and P.~Syverson, ``{AS}-awareness in {Tor} path selection,'' in
  \emph{Proceedings of the 16th ACM conference on Computer and communications
  security}.\hskip 1em plus 0.5em minus 0.4em\relax ACM, 2009, pp. 380--389.

\bibitem{feamster2004location}
N.~Feamster and R.~Dingledine, ``Location diversity in anonymity networks,'' in
  \emph{Proceedings of the 2004 ACM workshop on Privacy in the electronic
  society}.\hskip 1em plus 0.5em minus 0.4em\relax ACM, 2004, pp. 66--76.

\bibitem{johnson2013users}
A.~Johnson, C.~Wacek, R.~Jansen, M.~Sherr, and P.~Syverson, ``Users get routed:
  Traffic correlation on {Tor} by realistic adversaries,'' in \emph{Proceedings
  of the 2013 ACM SIGSAC conference on Computer \& communications
  security}.\hskip 1em plus 0.5em minus 0.4em\relax ACM, 2013, pp. 337--348.

\bibitem{sun2015raptor}
Y.~Sun, A.~Edmundson, L.~Vanbever, O.~Li, J.~Rexford, M.~Chiang, and P.~Mittal,
  ``Raptor: routing attacks on privacy in {Tor},'' in \emph{24th USENIX
  Security Symposium (USENIX Security 15)}, 2015, pp. 271--286.

\bibitem{akhoondi2012lastor}
M.~Akhoondi, C.~Yu, and H.~V. Madhyastha, ``{LasTor}: A low-latency {AS}-aware
  {Tor} client,'' in \emph{Security and Privacy (SP), 2012 IEEE Symposium
  on}.\hskip 1em plus 0.5em minus 0.4em\relax IEEE, 2012, pp. 476--490.

\bibitem{nithyanand2016measuring}
R.~Nithyanand, O.~Starov, A.~Zair, P.~Gill, and M.~Schapira, ``Measuring and
  mitigating {AS}-level adversaries against {Tor},'' in \emph{NDSS}, 2016.

\bibitem{topology}
``{CAIDA} {Internet} topology map,''
  \url{https://www.caida.org/research/topology/}.

\bibitem{torconsensus}
``Tor consensus,''
  \url{https://collector.torproject.org/recent/relay-descriptors/consensuses/}.

\bibitem{lad2007understanding}
M.~Lad, R.~Oliveira, B.~Zhang, and L.~Zhang, ``Understanding resiliency of
  {Internet} topology against prefix hijack attacks,'' in \emph{Dependable
  Systems and Networks, 2007. DSN'07. 37th Annual IEEE/IFIP International
  Conference on}.\hskip 1em plus 0.5em minus 0.4em\relax IEEE, 2007, pp.
  368--377.

\bibitem{backes2014nothing}
M.~Backes, A.~Kate, S.~Meiser, and E.~Mohammadi, ``(nothing else) {MATor}(s):
  Monitoring the anonymity of {Tor}'s path selection,'' in \emph{Proceedings of
  the 2014 ACM SIGSAC Conference on Computer and Communications
  Security}.\hskip 1em plus 0.5em minus 0.4em\relax ACM, 2014, pp. 513--524.

\bibitem{bgpstream}
``{BGP} stream.'' \url{http://bgpstream.caida.org/}.

\bibitem{murdoch2007sampled}
S.~J. Murdoch and P.~Zieli{\'n}ski, ``Sampled traffic analysis by
  {Internet}-exchange-level adversaries,'' in \emph{Privacy Enhancing
  Technologies}.\hskip 1em plus 0.5em minus 0.4em\relax Springer, 2007, pp.
  167--183.

\bibitem{indosat2014}
``Hijack event today by {Indosat},''
  \url{http://www.bgpmon.net/hijack-event-today-by-indosat/}.

\bibitem{barton2016denasa}
A.~Barton and M.~Wright, ``{DeNASA}: Destination-naive as-awareness in
  anonymous communications,'' \emph{Proceedings on Privacy Enhancing
  Technologies}, vol. 2016, no.~4, pp. 356--372, 2016.

\bibitem{tan2016data}
H.~Tan, M.~Sherr, and W.~Zhou, ``Data-plane defenses against routing attacks on
  {Tor},'' \emph{Proceedings on Privacy Enhancing Technologies}, vol.~4, pp.
  276--293, 2016.

\bibitem{indiahijack}
``Large scale {BGP} hijack out of {India}.''
  \url{http://www.bgpmon.net/large-scale-bgp-hijack-out-of-india/}.

\bibitem{syriahijack}
``{BGP} hijack incident by {Syrian} telecommunications establishment,''
  \url{http://www.bgpmon.net/bgp-hijack-incident-by-syrian-telecommunications-establishment/}.

\bibitem{many_hijack}
``Sprint, {Windstream}: Latest {ISPs} to hijack foreign networks,''
  \url{http://research.dyn.com/2014/09/latest-isps-to-hijack/}.

\bibitem{mada_hijack}
``On-going {BGP} hijack targets {Palestinian} {ISP},''
  \url{http://research.dyn.com/2015/01/going-bgp-attack-targets-palestinian-isp/}.

\bibitem{gao2001stable}
L.~Gao and J.~Rexford, ``Stable {Internet} routing without global
  coordination,'' \emph{IEEE/ACM Transactions on Networking (TON)}, vol.~9,
  no.~6, pp. 681--692, 2001.

\bibitem{gao2001inferring}
L.~Gao, ``On inferring autonomous system relationships in the internet,''
  \emph{IEEE/ACM Transactions on Networking (ToN)}, vol.~9, no.~6, pp.
  733--745, 2001.

\bibitem{ballani2007study}
H.~Ballani, P.~Francis, and X.~Zhang, ``A study of prefix hijacking and
  interception in the {Internet},'' in \emph{ACM SIGCOMM Computer Communication
  Review}, vol.~37, no.~4.\hskip 1em plus 0.5em minus 0.4em\relax ACM, 2007,
  pp. 265--276.

\bibitem{pilosov2008stealing}
A.~Pilosov and T.~Kapela, ``Stealing the internet: An internet-scale man in the
  middle attack,'' \emph{NANOG-44, Los Angeles, October}, pp. 12--15, 2008.

\bibitem{tille1996elimination}
Y.~Till{\'e}, ``An elimination procedure for unequal probability sampling
  without replacement,'' \emph{Biometrika}, vol.~83, no.~1, pp. 238--241, 1996.

\bibitem{maxmind}
``Maxmind {GeoLite} {ASN} database,''
  \url{http://dev.maxmind.com/geoip/legacy/geolite/}.

\bibitem{caida}
``The {IPv4} routed /24 topology dataset,''
  \url{http://www.caida.org/data/active/ipv4_routed_24_topology_dataset.xml}.

\bibitem{coifman1992entropy}
R.~R. Coifman and M.~V. Wickerhauser, ``Entropy-based algorithms for best basis
  selection,'' \emph{IEEE Transactions on information theory}, vol.~38, no.~2,
  pp. 713--718, 1992.

\bibitem{juen2012protecting}
J.~Juen, ``Protecting anonymity in the presence of autonomous system and
  {Internet} exchange level adversaries,'' 2012.

\bibitem{jansen2011shadow}
R.~Jansen and N.~Hopper, ``Shadow: Running {Tor} in a box for accurate and
  efficient experimentation,'' in \emph{NDSS}, 2011.

\bibitem{lad2006phas}
M.~Lad, D.~Massey, D.~Pei, Y.~Wu, B.~Zhang, and L.~Zhang, ``{PHAS}: A prefix
  hijack alert system.'' in \emph{Usenix Security}, 2006.

\bibitem{zhang2008ispy}
Z.~Zhang, Y.~Zhang, Y.~C. Hu, Z.~M. Mao, and R.~Bush, ``{iSpy}: detecting {IP}
  prefix hijacking on my own,'' in \emph{ACM SIGCOMM Computer Communication
  Review}, vol.~38, no.~4.\hskip 1em plus 0.5em minus 0.4em\relax ACM, 2008,
  pp. 327--338.

\bibitem{zheng2007light}
C.~Zheng, L.~Ji, D.~Pei, J.~Wang, and P.~Francis, ``A light-weight distributed
  scheme for detecting {IP} prefix hijacks in real-time,'' in \emph{ACM SIGCOMM
  Computer Communication Review}, vol.~37, no.~4.\hskip 1em plus 0.5em minus
  0.4em\relax ACM, 2007, pp. 277--288.

\bibitem{qiu2007detecting}
J.~Qiu, L.~Gao, S.~Ranjan, and A.~Nucci, ``Detecting bogus {BGP} route
  information: Going beyond prefix hijacking,'' in \emph{Security and Privacy
  in Communications Networks and the Workshops, 2007. SecureComm 2007. Third
  International Conference on}.\hskip 1em plus 0.5em minus 0.4em\relax IEEE,
  2007, pp. 381--390.

\bibitem{hu2007accurate}
X.~Hu and Z.~M. Mao, ``Accurate real-time identification of {IP} prefix
  hijacking,'' in \emph{Security and Privacy, 2007. SP'07. IEEE Symposium
  on}.\hskip 1em plus 0.5em minus 0.4em\relax IEEE, 2007, pp. 3--17.

\bibitem{shi2012detecting}
X.~Shi, Y.~Xiang, Z.~Wang, X.~Yin, and J.~Wu, ``Detecting prefix hijackings in
  the {Internet} with argus,'' in \emph{Proceedings of the 2012 ACM conference
  on Internet measurement conference}.\hskip 1em plus 0.5em minus 0.4em\relax
  ACM, 2012, pp. 15--28.

\bibitem{teamcymru}
``Team-{Cymru},'' \url{http://www.team-cymru.org/}.

\bibitem{malaysialeak}
``Massive route leak causes {Internet} slowdown.''
  \url{http://www.bgpmon.net/massive-route-leak-cause-internet-slowdown/}.

\bibitem{schlinker2014peering}
B.~Schlinker, K.~Zarifis, I.~Cunha, N.~Feamster, and E.~Katz-Bassett,
  ``{PEERING}: An {AS} for us,'' in \emph{Proceedings of the 13th ACM Workshop
  on Hot Topics in Networks}.\hskip 1em plus 0.5em minus 0.4em\relax ACM, 2014,
  p.~18.

\bibitem{juen2015defending}
J.~Juen, A.~Johnson, A.~Das, N.~Borisov, and M.~Caesar, ``Defending {Tor} from
  network adversaries: A case study of network path prediction,''
  \emph{Proceedings on Privacy Enhancing Technologies}, vol. 2015, no.~2, pp.
  171--187, 2015.

\end{thebibliography}
%
%
%

\end{document}